\documentclass[aip,pop,reprint]{revtex4-1}
\usepackage[T1]{fontenc}
\usepackage[utf8]{inputenc}
\usepackage{color}
\definecolor{shadecolor}{rgb}{0, 0.667969, 1}
\usepackage{units}
\usepackage{framed}
\usepackage{mathtools}
\usepackage{bm}
\usepackage{amsmath}
\usepackage{amssymb}
\usepackage{esint}
\usepackage[unicode=true,
 bookmarks=true,bookmarksnumbered=false,bookmarksopen=false,
 breaklinks=false,pdfborder={0 0 0},pdfborderstyle={},backref=false,colorlinks=true]
 {hyperref}
\hypersetup{pdftitle={A dispersion function for the regularized kappa distribution function},
 pdfauthor={Rudi Gaelzer},
 linkcolor=blue, citecolor=blue}

\makeatletter
\usepackage{mathptmx}
\usepackage{etoolbox}
\allowdisplaybreaks

\makeatletter
\def\@email#1#2{%
 \endgroup
 \patchcmd{\titleblock@produce}
  {\frontmatter@RRAPformat}
  {\frontmatter@RRAPformat{\produce@RRAP{*#1\href{mailto:#2}{#2}}}\frontmatter@RRAPformat}
  {}{}
}%

\makeatother

\begin{document}
\global\long\def\vuz{\hat{\bm{z}}}%
\global\long\def\erfc{\mathop{\mathrm{erfc}}\nolimits}%
\global\long\def\pFq#1#2{\prescript{\vphantom{#1}}{#1}{F}_{#2}^{\vphantom{#2}}}%

\title{A dispersion function for the regularized kappa distribution function}

\author{Rudi Gaelzer}
\affiliation{Instituto de Física, Universidade Federal do Rio Grande do Sul, CP 15051, 91501-970, Porto Alegre, RS, Brazil}
\email{rudi.gaelzer@ufrgs.br}

\author{Horst Fichtner}
\affiliation{Institut f\"ur Theoretische Physik IV, Ruhr-Universit\"at Bochum, 
Universit\"atsstrasse 150, 44780 Bochum, Germany}

\author{Klaus Scherer}
\affiliation{Institut f\"ur Theoretische Physik IV, Ruhr-Universit\"at Bochum, 
Universit\"atsstrasse 150, 44780 Bochum, Germany}


\begin{abstract}
In this work we define the plasma dispersion function for a suprathermal plasma described with a regularized kappa distribution. As is known from Maxwellian as well as (standard) kappa plasmas, the respective Fried-Conte and the modified plasma dispersion functions are  valuable tools for various analytical studies. For the latter is has been proven very useful to know about the mathematical properties, as analytical continuation, series expansions as well as asymptotic expressions. Given the growing popularity of the regularized kappa distribution, as indicated by its increasing number of applications to various problems related to suprathermal plasmas, we extend its theoretical treatment here by providing the corresponding plasma dispersion function along with various of its properties. 
\end{abstract}
\maketitle

\section{Introduction}
 
So-called plasma dispersion functions (PDFs) are very useful tools for the analytical formulation of dispersion relations describing waves in plasmas. These PDFs are also useful for the derivation of asymptotic expressions that are valuable for both analytical estimates and validation of numerical codes. 

The original PDF, nowadays also know as the Fried-Conte function, was introduced in Ref.\ \onlinecite{Fried-Conte-1961} as the Hilbert transform of the Gaussian, i.e.\ for an isotropic non-relativistic Maxwellian plasma. This was later generalized in Ref.\ \onlinecite{Godfrey-etal-1975} to the relativistic case employing the isotropic J\"uttner distribution\citep{Juettner-1911}. Further developments w.r.t.\ Maxwellian plasmas were the definition of generalized PDFs\citep{Percival-Robinson-1998}, relativistic quantum plasma dispersion functions\citep{Melrose-etal-2006}, the incomplete plasma dispersion function\citep{Peratt84/03,Baalrud-2013}, 
or bi-Maxwellian distributions\citep{Kunz-etal-2018}.

The concept of PDFs has also been generalized to other velocity distribution functions (VDFs), namely to the (standard) kappa distribution\citep{Summers-Thorne-1991,Summers-etal-1996, GaelzerZiebell14/12, GaelzerZiebell16/02, Gaelzer+16/06}, to the kappa-Maxwellian distribution\citep{Hellberg-Mace-2002}, to the Fermi-Dirac distribution\citep{Melrose-Mushtaq-2010, Mamedov-2012}, or the Druyvesteyn distribution\citep{Amemiya-2012}. See also Ref.\ \onlinecite{Xie-2013} for an emphasis on numerical evaluations of accordingly generalized PDFs.

The most often employed non-Maxwellian distribution is the standard kappa distribution (SKD), as numerous applications demonstrate\citep{Livadiotis17, LazarFichtner21}. Despite its success, there are drawbacks in the use of the SKD that comprise diverging velocity moments, a positive lower bound for kappa, and a non-extensive entropy, all of which entail unphysical features (see, e.g., the discussions in Refs.\ \onlinecite{LazarFichtner21}, \onlinecite{Scherer+17/12}, and \onlinecite{Fichtner-etal-2018}). Particularly with the motivation to find a formulation free of divergences, the (isotropic) \textit{regularized kappa distribution} (RKD) was introduced in Ref.\ \onlinecite{Scherer+17/12}. Following a subsequent demonstration that the RKD is indeed free of all undesired unphysical features\citep{Fichtner-etal-2018, Scherer-etal-2019} and that it can be 'derived' as a self-consistent solution for the quasi-equilibrium state between Langmuir fluctuations and suprathermal electrons\citep{Yoon-etal-2018}, the theory was extended to the anisotropic\citep{Scherer-etal-2019b} as well as the relativistic RKD\citep{HanThanh-etal-2022} and to a generalized RKD\citep{Scherer+20/07}.

Since then the RKD gained popularity as is evidenced by various applications to suprathermal plasmas, e.g., to collision-induced transport coefficients\citep{Husidic-etal-2022}, to fits of measured VDFs\citep{Scherer-etal-2022}, to various wave modes\citep{Liu-Chen-2022, Huo-Du-2023, Hammad-etal-2023}, to soliton structures\citep{Liu-2020, Lu-Liu-2021, Zhou-etal-2022, Li-Liu-2023}, to electron holes\citep{Haas-etal-2023}, to the Harris current sheet\citep{Hau-etal-2023}, or to the charging of dust\citep{Liu-2024}. Given this increasing interest into the RKD, it appears valuable to systematically extend the corresponding analytical theory. As outlined above, one highly useful tool is the PDF for a given VDF. Therefore, with the present paper we provide, in a similar fashion as Ref.\ \onlinecite{Fried-Conte-1961} for the Maxwellian and Ref.\ \onlinecite{Summers-Thorne-1991} for the SKD, the PDF for an (isotropic) RKD and study its mathematical properties such as the analytic continuation, power series as well as asymptotic expansions, all of which should be helpful for future analytical and numerical studies. 

The paper is structured as follows: after the theoretical formulation of the RKD PDF in section~\ref{theory}, its mathematical properties are analyzed in section~\ref{properties} 
and some plots of the PDF and its derivative are presented.
In section \ref{sec:RKAP1:Langmuir} the formalism developed here is applied to the description of the propagation and absorption of Langmuir waves in a pure RKD plasma.
All findings are briefly summarized in the concluding section~\ref{conclusions}.

\section{Theoretical formulation}\label{theory}

In this section, the basic formalism employed here will be introduced.

\subsection{The dielectric tensor and the dispersion equations}

\begin{subequations}
\label{eq:RKAP1:DT-magnetized-parallel-isotropic}

We will consider electrostatic/electromagnetic waves propagating parallelly
to an ambient (homogeneous) magnetic field $\bm{B}_{0}=B_{0}\vuz$
$\left(B_{0}>0\right)$\@. In this case,\citep{Brambilla98} the
only nonzero Cartesian components of the dielectric tensor are $\varepsilon_{xx}=\varepsilon_{1}$,
$\varepsilon_{xy}=-\varepsilon_{yx}=\varepsilon_{2}$ and $\varepsilon_{zz}=\varepsilon_{3}$,
which we will then write as 
\[
\varepsilon_{i}=\delta_{i1}+\delta_{i3}+\sum_{a}\chi_{i}^{(a)}\quad\left(i=1,2,3\right),
\]
where $\chi_{i}^{(a)}$ is the $i$-th component of the susceptibility
tensor associated with particle species/population $a=\left(e,i,\dots\right)$\@.
When the velocity distribution function is isotropic, \emph{i.e.},
$f_{a0}\left(\bm{v}\right)=f_{a0}\left(v\right)$, the susceptibility
components can be written as\citep{Brambilla98}

\begin{align}
\chi_{1}^{(a)}\left(k_{\parallel},\omega\right) & =-\frac{1}{2}\frac{\omega_{pa}^{2}}{\omega}\sum_{s=\pm1}\int\frac{f_{a0}\left(v\right)d^{3}v}{\omega-s\Omega_{a}-k_{\parallel}v_{\parallel}}\\
\chi_{2}^{(a)}\left(k_{\parallel},\omega\right) & =-\frac{i}{2}\frac{\omega_{pa}^{2}}{\omega}\sum_{s=\pm1}s\int\frac{f_{a0}\left(v\right)d^{3}v}{\omega-s\Omega_{a}-k_{\parallel}v_{\parallel}}\\
\chi_{3}^{(a)}\left(k_{\parallel},\omega\right) & =\frac{\omega_{pa}^{2}}{k_{\parallel}}\int d^{3}v\,\frac{\partial f_{a0}/\partial v_{\parallel}}{\omega-k_{\parallel}v_{\parallel}}.\label{eq:RKAP1:DTMPI-e3}
\end{align}
In (\ref{eq:RKAP1:DT-magnetized-parallel-isotropic}a-c), $v_{\parallel}$
and $v_{\perp}$ are, respectivelly, the parallel and perpendicular
(to $\bm{B}_{0}$) components of the particle velocity. Likewise,
$k_{\parallel}$ is the parallel component of the wavevector $\left(k_{\perp}=0\right)$
and $\omega$ is the wave angular frequency. Each particle species/population
is characterized by its mass $m_{a}$, charge $q_{a}$ and number
density $n_{a}$\@. Accordingly, $\omega_{pa}=\sqrt{4\pi n_{a}q_{a}^{2}/m_{a}}$
and $\Omega_{a}=q_{a}B_{0}/m_{a}c$ are the $a$-species (angular)
plasma and cyclotron frequencies, with $c$ being the vacuum light
speed. 
\end{subequations}

\begin{subequations}
\label{eq:RKAP1:Dispersion_equations}

The linearized Vlasov-Maxwell system of equations leads to the wave
equation for a magnetized plasma that, in the case of parallel propagation
and in the Fourier $\left(\bm{k},\omega\right)$ space factors into
the dispersion equation for longitudinal waves\citep{Brambilla98}
\begin{equation}
\varepsilon_{3}=1+\sum_{a}\frac{\omega_{pa}^{2}}{k_{\parallel}}\int d^{3}v\,\frac{\partial f_{a0}/\partial v_{\parallel}}{\omega-k_{\parallel}v_{\parallel}}=0,\label{eq:RKAP1:DE-electrostatic}
\end{equation}
and the dispersion equations for electromagnetic, circularly-polarized
waves 
\begin{equation}
N_{\parallel s}^{2}=\varepsilon_{1}-is\varepsilon_{2}=1-\sum_{a}\frac{\omega_{pa}^{2}}{\omega}\int\frac{f_{a0}\left(v\right)d^{3}v}{\omega-s\Omega_{a}-k_{\parallel}v_{\parallel}},
\end{equation}
where $s=\pm1$ and $\bm{N}=\bm{k}c/\omega$ is the index of refraction.
The equation for $s=+1$ corresponds to left-handed circularly polarized
waves $\left(N_{\parallel+}=N_{L}\right)$, whereas the equation for
$s=-1$ corresponds to right-handed waves $\left(N_{\parallel-}=N_{R}\right)$.
\end{subequations}

\subsection{A generalized form for the dispersion function}

From a given model for the VDF of a certain plasma species, the integrations
involved in the components of the susceptibility tensor (\ref{eq:RKAP1:DT-magnetized-parallel-isotropic}a-c)
are ideally expressed in terms of a single \emph{plasma dispersion
function} (PDF), which mathematically describes the contributions
of that species to the wave-particle interactions (\emph{i.e.}, Landau
or cyclotron damping) and the wave propagation characteristics (the
dispersion relations).
This objective is completely achieved when the VDF is the Maxwell-Boltzmann
distribution function
\[
f_{M}\left(v\right)=\frac{e^{-v^{2}/v_{T}^{2}}}{\pi^{3/2}v_{T}^{3}},
\]
where $v_{T}=\sqrt{2T/m}$ is the thermal speed, and $T$ is the (equilibrium)
temperature (in energy units)\@. As is well-known (see, \emph{e.g.},
Refs. \onlinecite{KrallTrivelpiece86,Brambilla98,Swanson08}), the
components in (\ref{eq:RKAP1:DT-magnetized-parallel-isotropic}) evaluated
with $f_{M}\left(v\right)$ can all be written in terms of the famous
Fried \& Conte dispersion function 
\begin{equation}
Z\left(\xi\right)=\frac{1}{\sqrt{\pi}}\int_{-\infty}^{\infty}\frac{e^{-t^{2}}dt}{t-\xi}\;\left(\xi_{i}>0\right),\label{eq:RKAP1:Fried-Conte-DF}
\end{equation}
and its derivative, both for parallel- and oblique-propagating waves.
The $Z$-function is entire in the complex plane $\xi=\xi_{r}+i\xi_{i}$
and so the analytic continuation of (\ref{eq:RKAP1:Fried-Conte-DF})
to the region $\xi_{i}\leqslant0$ is necessary. This is accomplished
via the Landau prescription.\citep{Brambilla98,KrallTrivelpiece86}

On the other hand, the task is only partially fulfilled in the case
of the standard kappa distribution function
\begin{equation}
f_{\kappa,w}\left(v\right)=\frac{1}{\left(\pi w^{2}\right)^{3/2}}\frac{\kappa^{-3/2}\Gamma\left(\sigma\right)}{\Gamma\left(\sigma-\nicefrac{3}{2}\right)}\left(1+\frac{v^{2}}{\kappa w^{2}}\right)^{-\sigma},\label{eq:RKAP1:SKD-general}
\end{equation}
where $\sigma=\kappa+\alpha$ $\left(\sigma>\nicefrac{3}{2}\right)$
and $w>0$ is a possibly $\kappa$-dependent parameter that is related
to the second moment of the VDF\@. The distribution $f_{\kappa,w}\left(v\right)$
was introduced by Refs. \onlinecite{GaelzerZiebell14/12,GaelzerZiebell16/02}
in order to provide a single expression that reproduces several different
models for the kappa distribution function employed in the literature.
From (\ref{eq:RKAP1:SKD-general}), one obtains the $\kappa$PDF\citep{GaelzerZiebell14/12,GaelzerZiebell16/02}
\begin{equation}
Z_{\kappa}^{\left(\alpha,\beta\right)}\left(\xi\right)=\frac{\kappa^{-\beta-1/2}}{\pi^{1/2}}\frac{\Gamma\left(\lambda-1\right)}{\Gamma\left(\sigma-\nicefrac{3}{2}\right)}\int_{-\infty}^{\infty}ds\frac{\left(1+s^{2}/\kappa\right)^{-\left(\lambda-1\right)}}{s-\xi},\label{eq:RKAP1:SKD-kPDF}
\end{equation}
again valid for $\xi_{i}>0$ and where $\lambda=\kappa+\alpha+\beta$
$\left(\lambda>1\right)$\@. In (\ref{eq:RKAP1:SKD-kPDF}), the parameter
$\beta\geqslant0$ makes it possible to write all the corresponding
expressions in (\ref{eq:RKAP1:DT-magnetized-parallel-isotropic})
and (\ref{eq:RKAP1:Dispersion_equations}) in terms of the same function
$Z_{\kappa}^{\left(\alpha,\beta\right)}\left(\xi\right)$.

However, as discussed by Refs. \onlinecite{GaelzerZiebell14/12,GaelzerZiebell16/02,Gaelzer+16/06},
for the case of oblique propagation, the dielectric tensor components
can not be simply written in terms of $Z_{\kappa}^{\left(\alpha,\beta\right)}\left(\xi\right)$\@.
The (mathematical) reason is because the distribution $f_{\kappa,w}\left(v\right)$
does not factor as the product of two functions, one containing only
$v_{\perp}$ and other containing only $v_{\parallel}$\@. Physically,
this stems from the fact that the $\kappa$VDF describes a (quasi-)
stationary turbulent state of a plasma where long-distance correlations
among the particles operate. Contrastingly, a plasma in thermodynamic
equilibrium is in a state where all correlations have vanished due
to Coulomb collisions and, hence, is described by the factorable Maxwell-Boltzmann
VDF $f_{M}\left(v\right)=f_{M\parallel}\left(v_{\parallel}\right)f_{M\perp}\left(v_{\perp}\right)$
(see discussions in Refs. \onlinecite{Livadiotis17,LazarFichtner21}).

The method we employ in this work to derive the dispersion function
for the regularized kappa VDF (to be introduced in sec. \ref{subsec:RKAP1:RkVDF-DF})
is based in a general expression for the component $\chi_{3}^{(a)}$
in (\ref{eq:RKAP1:DTMPI-e3})\@. For a given plasma species described
by the Maxwellian $f_{M}\left(v\right)$, the component becomes 
\[
\chi_{3,M}^{(a)}=-\frac{\omega_{pa}^{2}}{k_{\parallel}^{2}v_{Ta}^{2}}Z^{\prime}\left(\xi_{a}\right),
\]
where $\xi_{a}=\omega/k_{\parallel}v_{Ta}$ and $Z^{\prime}\left(\xi\right)=dZ/d\xi$\@.
On the other hand, if the species is suprathermal and described by
the SKD (\ref{eq:RKAP1:SKD-general}), then 
\[
\chi_{3,\mathrm{SKD}}^{(a)}=-\frac{\omega_{pa}^{2}}{k_{\parallel}^{2}w_{a}^{2}}Z_{\kappa}^{\left(\alpha,0\right)\prime}\left(\xi_{a}\right),
\]
where $\xi_{a}=\omega/k_{\parallel}w_{a}$ and $Z_{\kappa}^{(\alpha,\beta)\prime}\left(\xi\right)=dZ_{\kappa}^{(\alpha,\beta)}\left(\xi\right)/d\xi$.

Now, given that any gyrotropic VDF, \emph{i.e.}, with $f_{0}=f_{0}\left(v_{\perp},v_{\parallel}\right)$,
satisfies the identity 
\begin{align*}
\int d^{3}v\,\frac{\partial f_{0}/\partial v_{\parallel}}{v_{\parallel}-\chi} & =\frac{\partial\hphantom{\chi}}{\partial\chi}\tilde{I}_{3}\left(\chi\right), & \tilde{I}_{3}\left(\chi\right) & =\int d^{3}v\frac{f_{0}\left(v_{\perp},v_{\parallel}\right)}{v_{\parallel}-\chi},
\end{align*}
we can infer a generalized dispersion function, given simply by 
\[
Z^{(\mathrm{\mathrm{gen}})}\left(\xi\right)=\theta\tilde{I}_{3}\left(\xi\right),
\]
where $\xi=\omega/k_{\parallel}\theta$, and $\theta>0$ is a parameter
related to the second moment of $f_{0}\left(v_{\perp},v_{\parallel}\right)$\@.
In this case, the $\chi_{3,\mathrm{gen}}^{(a)}$ component can be
written as 
\[
\chi_{3,\mathrm{gen}}^{(a)}=-\frac{\omega_{pa}^{2}}{k_{\parallel}^{2}\theta_{a}^{2}}\frac{d\hphantom{\xi}}{d\xi_{a}}Z^{(\mathrm{\mathrm{gen}})}\left(\xi_{a}\right).
\]
Equivalently, by defining the new integration variables $v_{\parallel}=\theta s$
and $v_{\perp}=\theta x$, we can express the generalized function
as 
\begin{equation}
Z^{(\mathrm{\mathrm{gen}})}\left(\xi\right)=2\pi\theta^{3}\int_{-\infty}^{\infty}ds\,\int_{0}^{\infty}dx\,x\frac{f_{0}\left(x,s\right)}{s-\xi}.\label{eq:RKAP1:Z^(gen)-3}
\end{equation}

One can easily verify that when $f_{0}\left(v_{\perp},v_{\parallel}\right)=f_{M}\left(v\right)$,
with $\theta=v_{T}$, one obtains $Z^{(\mathrm{\mathrm{gen}})}\left(\xi\right)=Z\left(\xi\right)$,
given by (\ref{eq:RKAP1:Fried-Conte-DF})\@. Also, if $f_{0}\left(v_{\perp},v_{\parallel}\right)$
is the SKD (\ref{eq:RKAP1:SKD-general}) (with $\theta=w$), one obtains
$Z^{(\mathrm{\mathrm{gen}})}\left(\xi\right)=Z_{\kappa}^{(\alpha,0)}\left(\xi\right)$,
given by (\ref{eq:RKAP1:SKD-kPDF})\@. Hence, this simple method
will be used in the next section to derive an adequate expression
for the dispersion function of a plasma described by the regularized
kappa distribution function.

\subsection{The regularized distribution function and the associated dispersion
function\label{subsec:RKAP1:RkVDF-DF}}

As already discussed in the Introduction, the regularized $\kappa$
distribution function was introduced by Ref. \onlinecite{Scherer+17/12}
in order to correct some physical/mathematical inconsistencies posed
by the standard $\kappa$ distribution.
In this work, we will adopt the generalized $\kappa$-distribution
(GKD) defined by Ref. \onlinecite{Scherer+20/07}, written here as
\begin{equation}
f_{\kappa,\theta}^{(\eta,\zeta,\mu)}\left(v\right)=N_{\kappa,\theta}^{(\eta,\zeta,\mu)}\left(1+\frac{v^{2}}{\eta\theta^{2}}\right)^{-\zeta}e^{-\mu v^{2}/\theta^{2}},\label{eq:RKAP1:kRVDF-general}
\end{equation}
where $\eta=\eta\left(\kappa\right)$, $\zeta=\zeta\left(\kappa\right)$,
$\mu=\mu\left(\kappa\right)$ and $\theta= \theta\left(\kappa\right)$ are (possibly $\kappa$-dependent)
parameters and $N_{\kappa,\theta}^{(\eta,\zeta,\mu)}$ is the constant
that normalizes the distribution to unity, given by 
\begin{align*}
N_{\kappa,\theta}^{(\eta,\zeta,\mu)} & =\left[\left(\pi\eta\right)^{3/2}\theta^{3}U\left(\frac{3}{2},\frac{5}{2}-\zeta,\mu\eta\right)\right]^{-1}\\
 & =\left[\left(\frac{\pi}{\mu}\right)^{3/2}\left(\mu\eta\right)^{\zeta}\theta^{3}U\left(\zeta,\zeta-\frac{1}{2},\mu\eta\right)\right]^{-1},
\end{align*}
where $U\left(z\right)$ is one of the solutions of Kummer's equation,\citep{Daalhuis-NIST10a}
and is also known as the Tricomi function. The transition between
both expressions for $N_{\kappa,\theta}^{(\eta,\zeta,\mu)}$ is obtained
from Kummer's transformation 
\begin{equation}
U\left(a,b,z\right)=z^{1-b}U\left(a-b+1,2-b,z\right).\label{eq:RKAP1:Kummer-transformation}
\end{equation}

Distribution (\ref{eq:RKAP1:kRVDF-general}) reproduces several different
but related expressions for suprathermal distributions that were employed
in the literature in the last decades. In Ref. \onlinecite{Scherer+20/07},
the Reader can find tables showing how these different models are
reproduced by (\ref{eq:RKAP1:kRVDF-general}) via suitable combinations
of the parameters $\left\{ \eta,\zeta,\theta,\mu\right\} $\@. 
For our purposes, we observe that when $\mu=0$, the function $f_{\kappa,\theta}^{(\eta,\zeta,0)}\left(v\right)$
reduces to the standard kappa distribution. This can be shown by first
employing the limiting form for $\left|z\right|\ll1$,\citep{Daalhuis-NIST10a}
\[
U\left(a,b,z\right)=\frac{\Gamma\left(b-1\right)}{\Gamma\left(a\right)}z^{1-b}+\frac{\Gamma\left(1-b\right)}{\Gamma\left(a-b+1\right)}+\mathcal{O}\left(z^{2-b_{r}}\right),
\]
valid for $1\leqslant b_{r}<2$ $\left(b\neq1\right)$, where $\Gamma\left(z\right)$
is the Gamma function.\citep{AskeyRoy-NIST10} Then, if $\eta=\kappa$,
$\zeta=\sigma$ and $\theta=w$, we obtain $f_{\kappa,w}^{(\kappa,\sigma,0)}\left(v\right)=f_{\kappa,w}\left(v\right)$,
given by (\ref{eq:RKAP1:SKD-general}), which was already a generalized
definition.

Denoting now as $Z_{\kappa}^{(\eta,\zeta,\mu)}\left(\xi\right)$ the
convenient candidate for the dispersion function related to the generalized
$\kappa$-distribution, we insert (\ref{eq:RKAP1:kRVDF-general})
into (\ref{eq:RKAP1:Z^(gen)-3})\@. The integration in $x$ can be
identified with the integral representation of the Tricomi function\citep{Daalhuis-NIST10a}
\begin{equation}
U\left(a,b,z\right)=\frac{1}{\Gamma\left(a\right)}\int_{0}^{\infty}\mathrm{e}^{-zt}t^{a-1}(1+t)^{b-a-1}dt,\label{eq:RKAP1:Tricomi-Integ_rep-1}
\end{equation}
for $a_{r}>0$ and $\left|\arg\left(z\right)\right|<\pi/2$\@. 

\begin{subequations}
\label{eq:RKAP1:kRPDF-1}

Consequently, we obtain the following definition for the dispersion
function related to the GKD,
\begin{multline}
Z_{\kappa}^{(\eta,\zeta,\mu)}\left(\xi\right)=B_{\kappa}^{(\eta,\zeta,\mu)}\int_{-\infty}^{\infty}ds\,\frac{e^{-\mu s^{2}}}{s-\xi}\left(1+\frac{s^{2}}{\eta}\right)^{-\left(\zeta-1\right)}\\
\times U\left(1,2-\zeta,\mu\eta\left(1+\frac{s^{2}}{\eta}\right)\right),\label{eq:RKAP1:DF-1}
\end{multline}
once again valid for $\xi_{i}>0$ and where $B_{\kappa}^{(\eta,\zeta,\mu)}=\pi\eta\theta^{3}N_{\kappa,\theta}^{(\eta,\zeta,\mu)}=\left[\left(\pi\eta\right)^{1/2}U\left(\nicefrac{3}{2},\nicefrac{5}{2}-\zeta,\mu\eta\right)\right]^{-1}.$

A second form can be obtained from the first using the Kummer transformation
(\ref{eq:RKAP1:Kummer-transformation}), giving
\begin{multline}
Z_{\kappa}^{(\eta,\zeta,\mu)}\left(\xi\right)=\tilde{B}_{\kappa}^{(\eta,\zeta,\mu)}\int_{-\infty}^{\infty}ds\frac{e^{-\mu s^{2}}}{s-\xi}\\
\times U\left(\zeta,\zeta,\mu\eta\left(1+\frac{s^{2}}{\eta}\right)\right),\label{eq:RKAP1:DF-2}
\end{multline}
where $\tilde{B}_{\kappa}^{(\eta,\zeta,\mu)}=\left(\mu\eta\right)^{\zeta-1}B_{\kappa}^{(\eta,\zeta,\mu)}$.

Finally, a third (and simpler) expression is 
\begin{multline}
Z_{\kappa}^{(\eta,\zeta,\mu)}\left(\xi\right)=\tilde{B}_{\kappa}^{(\eta,\zeta,\mu)}\\
\times e^{\mu\eta}\int_{-\infty}^{\infty}ds\frac{\Gamma\left(1-\zeta,\mu\eta\left(1+s^{2}/\eta\right)\right)}{s-\xi},\label{eq:RKAP1:DF-3}
\end{multline}
\end{subequations}
which comes from the second, using the identity\citep{Daalhuis-NIST10a}
\begin{equation}
U\left(a,a,z\right)=e^{z}\Gamma\left(1-a,z\right),\label{eq:RKAP1:NIST10-13.6.6-1}
\end{equation}
where $\Gamma\left(\alpha,z\right)$ is the (upper) incomplete Gamma
function.\citep{Paris-NIST10}
This last expression permits us to characterize $Z_{\kappa}^{(\eta,\zeta,\mu)}\left(\xi\right)$  as the Hilbert transform of the incomplete Gamma function, albeit with a more complicated argument.
All versions of $Z_{\kappa}^{(\eta,\zeta,\mu)}\left(\xi\right)$ reduce
to $Z_{\kappa}^{\left(\alpha,0\right)}\left(\xi\right)$, given by
(\ref{eq:RKAP1:SKD-kPDF}), when $\eta=\kappa$, $\zeta=\sigma$ and
$\mu=0$\@.
The subsequent limit $\kappa\to\infty$ reduces the $\kappa$PDF to the Fried \& Conte function (\ref{eq:RKAP1:Fried-Conte-DF}).

\begin{subequations}
\label{eq:RKAP1:DT-RKD-parallel}

With the definitions (\ref{eq:RKAP1:kRPDF-1}a-c), the components
of the susceptibility tensor (\ref{eq:RKAP1:DT-magnetized-parallel-isotropic}a-c)
are finally given by 
\begin{align}
\chi_{1,\mathrm{GKD}}^{(a)}\left(k_{\parallel},\omega\right) & =\frac{1}{2}\frac{\omega_{pa}^{2}}{\omega k_{\parallel}\theta_{a}}\sum_{s=\pm1}Z_{\kappa}^{(\eta,\zeta,\mu)}\left(\xi_{sa}\right)\\
\chi_{2,\mathrm{GKD}}^{(a)}\left(k_{\parallel},\omega\right) & =\frac{i}{2}\frac{\omega_{pa}^{2}}{\omega k_{\parallel}\theta_{a}}\sum_{s=\pm1}sZ_{\kappa}^{(\eta,\zeta,\mu)}\left(\xi_{sa}\right)\\
\chi_{3,\mathrm{GKD}}^{(a)}\left(k_{\parallel},\omega\right) & =-\frac{\omega_{pa}^{2}}{k_{\parallel}^{2}\theta_{a}^{2}}Z_{\kappa}^{(\eta,\zeta,\mu)\prime}\left(\xi_{0a}\right),
\end{align}
where $\xi_{sa}=\left(\omega-s\Omega_{a}\right)/k_{\parallel}\theta_{a}$.
\end{subequations}

\section{Mathematical properties}\label{properties}

In this section, several properties, integral representations and
series expansions for the function $Z_{\kappa}^{(\eta,\zeta,\mu)}\left(\xi\right)$
and its derivative are derived.

\subsection{Analytic continuation}

\begin{subequations}
\label{eq:RKAP1:Analytic_continuation}

In case of direct numerical evaluation of the integrals in (\ref{eq:RKAP1:kRPDF-1}),
the analytic continuation for $\xi_{i}\leqslant0$ is needed. Straightforward
application of the Landau prescription\citep{KrallTrivelpiece86}
to (\ref{eq:RKAP1:DF-1}) determines that we can write the continued
function with a single expression as 

\begin{equation}
Z_{\kappa}^{(\eta,\zeta,\mu)}\left(\xi\right)=Z_{\kappa,NC}^{(\eta,\zeta,\mu)}\left(\xi\right)+2\pi i\epsilon Z_{\kappa,C}^{(\eta,\zeta,\mu)}\left(\xi\right),
\label{eq:RKAP1:Z_k-AC}
\end{equation}
where $Z_{\kappa,NC}^{(\eta,\zeta,\mu)}\left(\xi\right)$ is any of
the integrals in (\ref{eq:RKAP1:kRPDF-1}), 
\begin{equation}
\epsilon=\begin{cases}
0, & \xi_{i}>0\\
\nicefrac{1}{2}, & \xi_{i}=0\\
1, & \xi_{i}<0,
\end{cases}
\end{equation}
whereas 
\begin{widetext}
\begin{equation}
Z_{\kappa,C}^{(\eta,\zeta,\mu)}\left(\xi\right)=\left\{ \begin{aligned} & B_{\kappa}^{(\eta,\zeta,\mu)}e^{-\mu\xi^{2}}\left(1+\frac{\xi^{2}}{\eta}\right)^{-\left(\zeta-1\right)}U\left(1,2-\zeta,\mu\eta\left(1+\frac{\xi^{2}}{\eta}\right)\right)\\
 & \tilde{B}_{\kappa}^{(\eta,\zeta,\mu)}e^{-\mu\xi^{2}}U\left(\zeta,\zeta,\mu\eta\left(1+\frac{\xi^{2}}{\eta}\right)\right)\\
 & \tilde{B}_{\kappa}^{(\eta,\zeta,\mu)}e^{\mu\eta}\Gamma\left(1-\zeta,\mu\eta\left(1+\frac{\xi^{2}}{\eta}\right)\right),
\end{aligned}
\right.
\label{eq:RKAP1:Z_k,C^(e,z,m)}
\end{equation}
\end{widetext}
where each expression corresponds to the consecutive version of $Z_{\kappa,NC}^{(\eta,\zeta,\mu)}\left(\xi\right)$
in (\ref{eq:RKAP1:kRPDF-1}a-c). 
\end{subequations}

\subsection{Symmetry properties}

Some symmetry properties of $Z_{\kappa}^{(\eta,\zeta,\mu)}\left(\xi\right)$
can be obtained directly from the definition and its analytic continuation.

\begin{subequations}
\label{eq:RKAP1:Symmetry_properties}

Assuming initially that $\xi=\xi_{r}+i\xi_{i}$ with $\xi_{i}>0$\@.
Then $Z_{\kappa}^{(\eta,\zeta,\mu)}\left(\xi\right)$ is simply given
by (\ref{eq:RKAP1:kRPDF-1}), which can be written as 
\begin{multline*}
Z_{\kappa,NC}^{(\eta,\zeta,\mu)}\left(\xi\right)=\int_{-\infty}^{\infty}ds\,\frac{\left(s-\xi_{r}\right)z^{(\eta,\zeta,\mu)}\left(s\right)}{\left(s-\xi_{r}\right)^{2}+\xi_{i}^{2}}\\
+i\xi_{i}\int_{-\infty}^{\infty}ds\,\frac{z^{(\eta,\zeta,\mu)}\left(s\right)}{\left(s-\xi_{r}\right)^{2}+\xi_{i}^{2}},
\end{multline*}
where $z^{(\eta,\zeta,\mu)}\left(s\right)$ is a real function, such
that $z^{(\eta,\zeta,\mu)}\left(-s\right)=z^{(\eta,\zeta,\mu)}\left(s\right)$.
One can easily verify that 
\begin{align*}
Z_{\kappa,NC}^{(\eta,\zeta,\mu)}\left(-\xi\right) & =-Z_{\kappa,NC}^{(\eta,\zeta,\mu)}\left(\xi\right),\\
Z_{\kappa,NC}^{(\eta,\zeta,\mu)}\left(\xi^{*}\right) & =-Z_{\kappa,NC}^{(\eta,\zeta,\mu)*}\left(-\xi\right).
\end{align*}

On the other hand, when $\xi_{i}<0$, $Z_{\kappa}^{(\eta,\zeta,\mu)}\left(\xi\right)$
is given by (\ref{eq:RKAP1:Z_k-AC}), and one can directly verify
that $Z_{\kappa,C}^{(\eta,\zeta,\mu)}\left(-\xi\right)=Z_{\kappa,C}^{(\eta,\zeta,\mu)}\left(\xi\right)$\@.
Therefore, we have one of the usual symmetry properties of a PDF,
\begin{equation}
Z_{\kappa}^{(\eta,\zeta,\mu)}\left(\xi\right)+Z_{\kappa}^{(\eta,\zeta,\mu)}\left(-\xi\right)=2\pi iZ_{\kappa,C}^{(\eta,\zeta,\mu)}\left(\xi\right),\label{eq:RKAP1:SP-1}
\end{equation}
which also stands when $\xi_{i}=0$.

Now, all functions $f_{1}\left(z\right)=z^{-\left(\zeta-1\right)}$,
$f_{2}\left(z\right)=U\left(a,b,z\right)$, and $f_{3}\left(z\right)=\Gamma\left(1-\zeta,z\right)$
in (\ref{eq:RKAP1:Z_k,C^(e,z,m)}) have branch cuts in the complex
$\xi$ plane, running along $\left(-i\infty,-i\sqrt{\eta}\right]\cup\left[i\sqrt{\eta},+i\infty\right)$
when $\zeta$ is not integer\@.\citep{Daalhuis-NIST10a,Paris-NIST10}
Hence, $Z_{\kappa,C}^{(\eta,\zeta,\mu)}\left(\xi^{*}\right)=Z_{\kappa,C}^{(\eta,\zeta,\mu)*}\left(\xi\right)$\@.
Therefore, we have another usual property for dispersion functions,
\begin{equation}
Z_{\kappa}^{(\eta,\zeta,\mu)}\left(\xi^{*}\right)=-Z_{\kappa}^{(\eta,\zeta,\mu)*}\left(-\xi\right).\label{eq:RKAP1:SP-2}
\end{equation}

Finally, combining (\ref{eq:RKAP1:SP-1}) with (\ref{eq:RKAP1:SP-2}),
we have the third symmetry property 
\begin{equation}
Z_{\kappa}^{(\eta,\zeta,\mu)}\left(\xi^{*}\right)=Z_{\kappa}^{(\eta,\zeta,\mu)*}\left(\xi\right)+2\pi iZ_{\kappa,C}^{(\eta,\zeta,\mu)}\left(\xi^{*}\right),\label{eq:RKAP1:SP-3}
\end{equation}
which is useful for the computer implementation of $Z_{\kappa}^{(\eta,\zeta,\mu)}\left(\xi\right)$.
\end{subequations}

We also observe that due to the fact that the analytic continuation
for $\xi_{i}<0$ is determined by $Z_{\kappa,C}^{(\eta,\zeta,\mu)}\left(\xi\right)$,
the regularized $\kappa$ PDF has a branch cut running along the imaginary
$\xi$ axis, in the interval $\left(-i\infty,-i\sqrt{\eta}\right]$.

\subsection{Value at $\xi=0$}

In this case, we return, for instance, to (\ref{eq:RKAP1:DF-3}) and
write
\begin{multline*}
Z_{\kappa}^{(\eta,\zeta,\mu)}\left(0\right)=\tilde{B}_{\kappa}^{(\eta,\zeta,\mu)}\\
\times e^{\mu\eta}\int_{-\infty}^{\infty}\frac{ds}{s}\Gamma\left(1-\zeta,\mu\eta\left(1+\frac{s^{2}}{\eta}\right)\right).
\end{multline*}
Using Plemelj's formula and Landau prescription,

\[
\lim_{\epsilon\to0}\int\frac{f\left(x\right)}{x-\left(x_{0}+i\epsilon\right)}dx=\fint\frac{f\left(x\right)}{x-x_{0}}dx+i\pi f\left(x_{0}\right),
\]
where $\fint$ denotes the principal part of the integral, the latter
is zero, because the integrand is an odd function of $s$\@. 

Therefore, 
\begin{equation}
\begin{aligned}Z_{\kappa}^{(\eta,\zeta,\mu)}\left(0\right) & =i\frac{\left(\pi\mu\right)^{1/2}e^{\mu\eta}\Gamma\left(1-\zeta,\mu\eta\right)}{U\left(\zeta,\zeta-\frac{1}{2},\mu\eta\right)}\\
 & =i\left(\pi\mu\right)^{1/2}\frac{U\left(\zeta,\zeta,\mu\eta\right)}{U\left(\zeta,\zeta-\frac{1}{2},\mu\eta\right)},
\end{aligned}
\label{eq:RKAP1:DF3-value-xi0}
\end{equation}
where the second expression was obtained from (\ref{eq:RKAP1:NIST10-13.6.6-1}).

\subsection{Integral representation}


An useful integral representation for $Z_{\kappa}^{(\eta,\zeta,\mu)}\left(\xi\right)$
can be obtained from the identity 
\begin{equation}
\int_{0}^{\infty}\beta^{-\delta}e^{-\gamma\beta}d\beta=\Gamma\left(1-\delta\right)\gamma^{\delta-1}\quad\left(\gamma>0\right),\label{eq:RKAP1:Superstatistics-ident-1}
\end{equation}
which will provide a representation in terms of the Fried \& Conte
function. From this identity, we can write
\[
\left(1+\frac{v^{2}}{\eta\theta^{2}}\right)^{-\zeta}=\frac{1}{\Gamma\left(\zeta\right)}\int_{0}^{\infty}d\beta\,\beta^{\zeta-1}e^{-\beta}e^{-\beta v^{2}/\eta\theta^{2}},
\]
and then write (\ref{eq:RKAP1:kRVDF-general}) as 
\begin{equation}
f_{\kappa,\theta}^{(\eta,\zeta,\mu)}\left(v\right)=\frac{N_{\kappa,\theta}^{(\eta,\zeta,\mu)}}{\Gamma\left(\zeta\right)}e^{-\mu v^{2}/\theta^{2}}\int_{0}^{\infty}d\beta\,\beta^{\zeta-1}e^{-\left(1+v^{2}/\eta\theta^{2}\right)\beta}.\label{eq:RKAP1:kRVDF-general-superstatistics}
\end{equation}

Inserting (\ref{eq:RKAP1:kRVDF-general-superstatistics}) in (\ref{eq:RKAP1:Z^(gen)-3})
and performing the integration in $x$, we obtain 
\begin{multline*}
Z_{\kappa}^{(\eta,\zeta,\mu)}\left(\xi\right)=\pi\theta^{3}\frac{N_{\kappa,\theta}^{(\eta,\zeta,\mu)}}{\Gamma\left(\zeta\right)}\int_{0}^{\infty}d\beta\,\frac{\beta^{\zeta-1}e^{-\beta}}{\mu+\beta/\eta}\\
\times\int_{-\infty}^{\infty}ds\,\frac{e^{-s^{2}}}{s-\xi_{\mathrm{eff}}\left(\beta\right)},
\end{multline*}
where 
\begin{gather*}
\xi_{\mathrm{eff}}\left(\beta\right)=\xi\sqrt{\mu+\frac{\beta}{\eta}}.
\end{gather*}
The last integral can be readily identified with the Fried \& Conte
function (\ref{eq:RKAP1:Fried-Conte-DF}), resulting then 
\begin{equation}
Z_{\kappa}^{(\eta,\zeta,\mu)}\left(\xi\right)=\frac{\sqrt{\pi}B_{\kappa}^{(\eta,\zeta,\mu)}}{\eta\Gamma\left(\zeta\right)}\int_{0}^{\infty}d\beta\,\frac{\beta^{\zeta-1}e^{-\beta}}{\mu+\beta/\eta}Z\left(\xi_{\mathrm{eff}}\right).\label{eq:RKAP1:DF-Superstatistics}
\end{equation}
This integral representation is useful to evaluate $Z_{\kappa}^{(\eta,\zeta,\mu)}\left(\xi\right)$
when $\xi_{i}\geqslant0$\@. For the case $\xi_{i}<0$, one can employ
the symmetry condition (\ref{eq:RKAP1:SP-3}).

From (\ref{eq:RKAP1:DF-Superstatistics}) one can also obtain different
representations for $Z_{\kappa}^{(\eta,\zeta,\mu)}\left(\xi\right)$
as series expansions, that will converge in different regions of the
complex $\xi$ plane.


\subsection{Power series expansions\label{subsec:RKAP1:Power_series_expansion}}

We will start from (\ref{eq:RKAP1:DF-Superstatistics}) and derive
two power series expansions for $Z_{\kappa}^{(\eta,\zeta,\mu)}\left(\xi\right)$\@.
As is well-known (see, \emph{e.g.}, Ref. \onlinecite{Swanson08}),
the $Z$-function has the following representation:
\[
Z\left(\xi\right)=i\sqrt{\pi}\mathrm{e}^{-\xi^{2}}\erfc\left(-i\xi\right)=i\sqrt{\pi}w\left(\xi\right),
\]
with $\erfc\left(z\right)$ being the complementary error function.\citep{Temme-NIST10-1}
The function $w\left(\xi\right)$ is also known as the Faddeeva function,
and both are entire functions in the $\xi$ plane.

\begin{subequations}
\label{eq:RKAP1:Z-function-expansions}

We start from the power series expansions of the Faddeeva function\citep{Temme-NIST10-1}
\[
w\left(z\right)=\sum_{n=0}^{\infty}\frac{\left(iz\right)^{n}}{\Gamma\left(1+\frac{n}{2}\right)},
\]
which has an infinite radius of convergence. Separating into two series,
with even and odd powers, one obtains\citep{Swanson08} 
\begin{equation}
Z\left(\xi\right)=i\sqrt{\pi}e^{-\xi^{2}}-2\xi\sum_{n=0}^{\infty}\frac{\left(-\xi^{2}\right)^{n}}{\left(\nicefrac{3}{2}\right)_{n}},\label{eq:RKAP1:ZFE-1}
\end{equation}
where $\left(\alpha\right)_{n}=\Gamma\left(\alpha+n\right)/\Gamma\left(\alpha\right)$
is the Pochhammer symbol. This is a known power series for the Fried
\& Conte function, with infinite radius of convergence.

Another series can be obtained from\citep{Temme-NIST10-1}
\[
\mathrm{erfc}\left(z\right)=1-\frac{2}{\sqrt{\pi}}\sum_{n=0}^{\infty}\frac{\left(-\right)^{n}z^{2n+1}}{\left(2n+1\right)n!}\;\left(\left|z\right|<\infty\right),
\]
in which case we have
\begin{equation}
Z\left(\xi\right)=\mathrm{e}^{-\xi^{2}}\left[i\sqrt{\pi}-2\sum_{n=0}^{\infty}\frac{\xi^{2n+1}}{\left(2n+1\right)n!}\right],\label{eq:RKAP1:ZFE-2}
\end{equation}
which also has an infinite radius of convergence.
\end{subequations}

So, inserting the first expansion (\ref{eq:RKAP1:ZFE-1}) into (\ref{eq:RKAP1:DF-Superstatistics}),
we obtain
\[
Z_{\kappa}^{(\eta,\zeta,\mu)}\left(\xi\right)=\frac{\sqrt{\pi}B_{\kappa}^{(\eta,\zeta,\mu)}}{\eta\Gamma\left(\zeta\right)}\left[i\sqrt{\pi}e^{-\mu\xi^{2}}I_{1}-2\xi\sum_{n=0}^{\infty}\frac{\left(-\xi^{2}\right)^{n}}{\left(\nicefrac{3}{2}\right)_{n}}I_{2}^{(n)}\right],
\]
where
\begin{align*}
I_{1} & =\int_{0}^{\infty}d\beta\,\beta^{\zeta-1}\left(\mu+\frac{\beta}{\eta}\right)^{-1}e^{-\left(1+\xi^{2}/\eta\right)\beta}\\
I_{2}^{(n)} & =\int_{0}^{\infty}d\beta\,\beta^{\zeta-1}\left(\mu+\frac{\beta}{\eta}\right)^{n-1/2}e^{-\beta}.
\end{align*}
But, according to (\ref{eq:RKAP1:Tricomi-Integ_rep-1}),
\begin{align*}
I_{1} & =\eta\left(\mu\eta\right)^{\zeta-1}\Gamma\left(\zeta\right)U\left(\zeta,\zeta,\mu\eta\left(1+\frac{\xi^{2}}{\eta}\right)\right)\\
I_{2}^{(n)} & =\mu^{n-1/2}\left(\mu\eta\right)^{\zeta}\Gamma\left(\zeta\right)U\left(\zeta,\zeta+n+\frac{1}{2},\mu\eta\right).
\end{align*}
Therefore, we obtain the first series expansion 
\begin{multline}
Z_{\kappa}^{(\eta,\zeta,\mu)}\left(\xi\right)=-2\sqrt{\pi}B_{\kappa}^{(\eta,\zeta,\mu)}\\
\times\frac{\xi}{\sqrt{\eta}}\sum_{\ell=0}^{\infty}\frac{y_{\ell}\left(\zeta,\zeta+\nicefrac{1}{2};\mu\eta\right)}{\left(\nicefrac{3}{2}\right)_{\ell}}\left(-\frac{\xi^{2}}{\eta}\right)^{\ell}\\
+i\pi\tilde{B}_{\kappa}^{(\eta,\zeta,\mu)}e^{-\mu\xi^{2}}U\left(\zeta,\zeta,\mu\eta\left(1+\frac{\xi^{2}}{\eta}\right)\right),\label{eq:RKAP1:kRPDF-4a}
\end{multline}
where 
\begin{equation}
y_{\ell}\left(a,b;z\right)=z^{b-1+\ell}U\left(a,b+\ell,z\right).\label{eq:RKAP1:yl-def}
\end{equation}

Since the Tricomi function satisfies the recurrence relation\citep{Daalhuis-NIST10a}
\begin{multline*}
\left(b-a-1\right)U\left(a,b-1,z\right)+\left(1-b-z\right)U\left(a,b,z\right)\\
+zU\left(a,b+1,z\right)=0,
\end{multline*}
the function $y_{\ell}\left(a,b;z\right)$ satisfies
\begin{multline*}
y_{\ell+1}\left(a,b;z\right)-\left(b-1+z+\ell\right)y_{\ell}\left(a,b;z\right)\\
+\left(b-a-1+\ell\right)zy_{\ell-1}\left(a,b;z\right)=0.
\end{multline*}

Inserting now the second expansion (\ref{eq:RKAP1:ZFE-2}) into (\ref{eq:RKAP1:DF-Superstatistics}),
we obtain
\[
Z_{\kappa}^{(\eta,\zeta,\mu)}\left(\xi\right)=\frac{\sqrt{\pi}B_{\kappa}^{(\eta,\zeta,\mu)}}{\eta\Gamma\left(\zeta\right)}\mathrm{e}^{-\mu\xi^{2}}\left[i\sqrt{\pi}I_{1}-2\sum_{n=0}^{\infty}\frac{\xi^{2n+1}I_{3}}{\left(2n+1\right)n!}\right],
\]
where 
\begin{gather*}
I_{3}=\int_{0}^{\infty}d\beta\,\beta^{\zeta-1}\left(\mu+\frac{\beta}{\eta}\right)^{n-1/2}e^{-\left(1+\xi^{2}/\eta\right)\beta},\\
I_{3}=\mu^{n-1/2}\left(\mu\eta\right)^{\zeta}\Gamma\left(\zeta\right)U\left(\zeta,\zeta+n+\frac{1}{2},\mu\eta\left(1+\frac{\xi^{2}}{\eta}\right)\right),
\end{gather*}
using again (\ref{eq:RKAP1:Tricomi-Integ_rep-1}).
Hence, we obtain the second expansion
\begin{multline}
Z_{\kappa}^{(\eta,\zeta,\mu)}\left(\xi\right)=-2\sqrt{\pi}B_{\kappa}^{(\eta,\zeta,\mu)}\mathrm{e}^{-\mu\xi^{2}}\frac{\xi}{\sqrt{\eta}}\left(1+\frac{\xi^{2}}{\eta}\right)^{1/2-\zeta}\\
\times\sum_{n=0}^{\infty}\frac{y_{n}\left(\zeta,\zeta+\frac{1}{2};\mu\eta\left(1+\xi^{2}/\eta\right)\right)}{\left(2n+1\right)n!}\left(\frac{\xi^{2}/\eta}{1+\xi^{2}/\eta}\right)^{n}\\
+i\pi\tilde{B}_{\kappa}^{(\eta,\zeta,\mu)}e^{\mu\eta}\Gamma\left(1-\zeta,\mu\eta\left(1+\frac{\xi^{2}}{\eta}\right)\right),\label{eq:RKAP1:kRPDF-5}
\end{multline}
where we have used (\ref{eq:RKAP1:NIST10-13.6.6-1}) and (\ref{eq:RKAP1:yl-def}).

\subsection{Asymptotic expansion}

Given the geometric progression 
\[
\sum_{m=0}^{N}q^{m}=\frac{1-q^{N+1}}{1-q}\:\left(q\neq1\right),
\]
we can write (\ref{eq:RKAP1:Fried-Conte-DF}) as 
\begin{align}
Z\left(\xi\right) & =-\frac{1}{\xi}\sum_{m=0}^{\left\lfloor N/2\right\rfloor }\frac{\left(\nicefrac{1}{2}\right)_{m}}{\xi^{2m}}+R_{N}\left(\xi\right)\;\left(\xi_{i}>0\right),\label{eq:RKAP1:Z-function-asymptotic}\\
R_{N}\left(\xi\right) & =-\frac{1}{\sqrt{\pi}}\frac{1}{\xi^{N+2}}\int_{-\infty}^{\infty}dt\,\frac{t^{N+1}\mathrm{e}^{-t^{2}}}{1-t/\xi},\nonumber 
\end{align}
where $\left\lfloor x\right\rfloor $ is the floor function and $R_{N}\left(\xi\right)$
is the remainder of the finite expansion.
This expansion can be inserted into (\ref{eq:RKAP1:DF-Superstatistics}),
thus rendering the asymptotic expansion of $Z_{\kappa}^{(\eta,\zeta,\mu)}\left(\xi\right)$
when $\left|\xi/\sqrt{\eta}\right|\gg1$\@. By doing that, one obtains
\begin{align*}
Z_{\kappa}^{(\eta,\zeta,\mu)}\left(\xi\right) & \sim-\frac{\sqrt{\pi}B_{\kappa}^{(\eta,\zeta,\mu)}}{\eta\Gamma\left(\zeta\right)}\sum_{m=0}\frac{\left(\nicefrac{1}{2}\right)_{m}}{\xi^{2m+1}}I_{2}^{(-m-1)},
\end{align*}
where $I_{2}^{(n)}$ is the integral defined in section \ref{subsec:RKAP1:Power_series_expansion}.
Hence, the asymptotic expansion of $Z_{\kappa}^{(\eta,\zeta,\mu)}\left(\xi\right)$,
taking into account the Landau contour for $\xi_{i}\leqslant0$, can
be written as 
\begin{multline}
Z_{\kappa}^{(\eta,\zeta,\mu)}\left(\xi\right)\sim-\sqrt{\pi}B_{\kappa}^{(\eta,\zeta,\mu)}\frac{\sqrt{\eta}}{\xi}\\
\times\sum_{m=0}\left(\frac{1}{2}\right)_{m}U\left(m+\frac{3}{2},m+\frac{5}{2}-\zeta,\mu\eta\right)\left(\frac{\eta}{\xi^{2}}\right)^{m}\\
+2\pi i\epsilon Z_{\kappa,C}^{(\eta,\zeta,\mu)}\left(\xi\right),\label{eq:RKAP1:kRPDF-asymptotic}
\end{multline}
where the Kummer transformation (\ref{eq:RKAP1:Kummer-transformation})
was also employed.

\subsection{The derivative of $Z_{\kappa}^{(\eta,\zeta,\mu)}\left(\xi\right)$
and its properties}

The derivative $Z_{\kappa}^{(\eta,\zeta,\mu)\prime}\left(\xi\right)=dZ_{\kappa}^{(\eta,\zeta,\mu)}/d\xi$
is also an important function for the evaluation of the dielectric
tensor. Therefore, its evaluation and mathematical properties need
to be addressed as well.

We can start by deriving directly (\ref{eq:RKAP1:DF-3}), for instance.
Since\citep{Paris-NIST10}
\begin{align*}
\frac{d\hphantom{z}}{dz}\Gamma\left(\alpha,z\right) & =-z^{\alpha-1}e^{-z}\\
\Gamma\left(\alpha,z\right) & =z^{\alpha-1}e^{-z}\left(\sum_{m=0}^{N-1}\frac{u_{m}}{z^{m}}+\mathcal{O}\left(z^{-N}\right)\right),
\end{align*}
where $u_{m}=\left(-\right)^{m}\left(1-\alpha\right)_{m}$, the derivation
followed by an integration by parts provides the intermediate result
\begin{equation}
Z_{\kappa}^{(\eta,\zeta,\mu)\prime}\left(\xi\right)=-\frac{2}{\eta}B_{\kappa}^{(\eta,\zeta,\mu)}\int_{-\infty}^{\infty}du\frac{ue^{-\mu u^{2}}}{u-\xi}\left(1+\frac{u^{2}}{\eta}\right)^{-\zeta}.\label{eq:RKAP1:kRPDF-deriv-1}
\end{equation}

\subsubsection{Value at $\xi=0$}

Result (\ref{eq:RKAP1:kRPDF-deriv-1}) can be used to evaluate $Z_{\kappa}^{(\eta,\zeta,\mu)\prime}\left(0\right)$,
which then follows directly from the identity (\ref{eq:RKAP1:Tricomi-Integ_rep-1}),
resulting 
\begin{equation}
Z_{\kappa}^{(\eta,\zeta,\mu)\prime}\left(0\right)=-2\mu\frac{U\left(\zeta,\zeta+\frac{1}{2},\mu\eta\right)}{U\left(\zeta,\zeta-\frac{1}{2},\mu\eta\right)},\label{eq:RKAP1:kRPDF-deriv-xi0}
\end{equation}
after using also (\ref{eq:RKAP1:Kummer-transformation}).

\subsubsection{Formula for the derivative}

\begin{subequations}
\label{eq:RKAP1:kRPDF-deriv-2}

Returning to (\ref{eq:RKAP1:kRPDF-deriv-1}) and completing the denominator,
we get 
\begin{equation}
Z_{\kappa}^{(\eta,\zeta,\mu)\prime}\left(\xi\right)=Z_{\kappa}^{(\eta,\zeta,\mu)\prime}\left(0\right)-2\xi D_{\kappa}^{(\eta,\zeta,\mu)}\left(\xi\right),\label{eq:RKAP1:kRPDF-D2-Z}
\end{equation}
where 
\begin{equation}
D_{\kappa}^{(\eta,\zeta,\mu)}\left(\xi\right)=\tilde{D}_{\kappa}^{(\eta,\zeta,\mu)}\int_{-\infty}^{\infty}du\frac{e^{-\mu u^{2}}}{u-\xi}\left(1+\frac{u^{2}}{\eta}\right)^{-\zeta},\label{eq:RKAP1:kRPDF-D2-D}
\end{equation}
and $\tilde{D}_{\kappa}^{(\eta,\zeta,\mu)}=B_{\kappa}^{(\eta,\zeta,\mu)}/\eta$.
\end{subequations}

We notice that, contrary to the $Z\left(\xi\right)$ and $Z_{\kappa}^{\left(\alpha,\beta\right)}\left(\xi\right)$
functions, the derivative of $Z_{\kappa}^{(\eta,\zeta,\mu)}\left(\xi\right)$
is not expressed as a recurrence relation involving itself. Instead,
it is given by the new function $D_{\kappa}^{(\eta,\zeta,\mu)}\left(\xi\right)$\@.

\subsubsection{Integral representations}

The integral representation of $Z_{\kappa}^{(\eta,\zeta,\mu)\prime}\left(\xi\right)$
is obtained in a straightforward way via direct differentiation of
(\ref{eq:RKAP1:DF-Superstatistics}) with respect to $\xi$, resulting
in 
\begin{equation}
Z_{\kappa}^{(\eta,\zeta,\mu)\prime}\left(\xi\right)=\frac{\sqrt{\pi}B_{\kappa}^{(\eta,\zeta,\mu)}}{\eta\Gamma\left(\zeta\right)}\int_{0}^{\infty}d\beta\,\frac{\beta^{\zeta-1}e^{-\beta}}{\sqrt{\mu+\beta/\eta}}Z^{\prime}\left(\xi_{\mathrm{eff}}\right).\label{eq:RKAP1:DF-deriv-Superstatistics}
\end{equation}

On the other hand, an integral representation for the function $D_{\kappa}^{(\eta,\zeta,\mu)}\left(\xi\right)$
is also possible, if we first consider the identity (\ref{eq:RKAP1:Superstatistics-ident-1})
as 
\[
\left(1+\frac{u^{2}}{\eta}\right)^{-\zeta}=\frac{1}{\Gamma\left(\zeta\right)}\int_{0}^{\infty}d\beta\,\beta^{\zeta-1}e^{-\left(1+u^{2}/\eta\right)\beta}.
\]
In this case, we can write (\ref{eq:RKAP1:kRPDF-D2-D}) as
\begin{equation}
D_{\kappa}^{(\eta,\zeta,\mu)}\left(\xi\right)=\frac{\sqrt{\pi}\tilde{D}_{\kappa}^{(\eta,\zeta,\mu)}}{\Gamma\left(\zeta\right)}\int_{0}^{\infty}d\beta\,\beta^{\zeta-1}e^{-\beta}Z\left(\xi_{\mathrm{eff}}\right).\label{eq:RKAP1:kRPDF-D-Superstatistics}
\end{equation}

\subsubsection{Power series expansions}

Series expansions for the functions $Z_{\kappa}^{(\eta,\zeta,\mu)\prime}\left(\xi\right)$
and $D_{\kappa}^{(\eta,\zeta,\mu)}\left(\xi\right)$ can be obtained
directly from the integral representations (\ref{eq:RKAP1:DF-deriv-Superstatistics})
and (\ref{eq:RKAP1:kRPDF-D-Superstatistics}), by introducing the
corresponding series expansion of the $Z-$function and its derivative,
the latter being
\[
Z^{\prime}\left(\zeta\right)=-2i\sqrt{\pi}\zeta\mathrm{e}^{-\zeta^{2}}-2\sum_{n=0}^{\infty}\frac{\left(-\zeta^{2}\right)^{n}}{\left(\nicefrac{1}{2}\right)_{n}}.
\]

In this case, one obtains
\begin{multline}
Z_{\kappa}^{(\eta,\zeta,\mu)\prime}\left(\xi\right)=-2\sqrt{\pi}\tilde{D}_{\kappa}^{(\eta,\zeta,\mu)}\\
\times\left[\sqrt{\eta}\sum_{\ell=0}^{\infty}\frac{y_{\ell}\left(\zeta,\zeta+\nicefrac{1}{2};\mu\eta\right)}{\left(\nicefrac{1}{2}\right)_{\ell}}\left(-\frac{\xi^{2}}{\eta}\right)^{\ell}\right.\\
\left.+i\sqrt{\pi}\xi\left(1+\frac{\xi^{2}}{\eta}\right)^{-\zeta}e^{-\mu\xi^{2}}\right],\label{eq:RKAP1:kRPDF-deriv-3}
\end{multline}
and 
\begin{multline}
D_{\kappa}^{(\eta,\zeta,\mu)}\left(\xi\right)=-\sqrt{\pi}\tilde{D}_{\kappa}^{(\eta,\zeta,\mu)}\\
\times\left[2\frac{\xi}{\sqrt{\eta}}\sum_{\ell=0}^{\infty}\frac{y_{\ell}\left(\zeta,\zeta+\nicefrac{3}{2};\mu\eta\right)}{\left(\nicefrac{3}{2}\right)_{\ell}}\left(-\frac{\xi^{2}}{\eta}\right)^{\ell}\right.\\
\left.-i\sqrt{\pi}\left(1+\frac{\xi^{2}}{\eta}\right)^{-\zeta}e^{-\mu\xi^{2}}\right].\label{eq:RKAP1:RKD-series-1}
\end{multline}
Expansion (\ref{eq:RKAP1:RKD-series-1}) for $D_{\kappa}^{(\eta,\zeta,\mu)}\left(\xi\right)$
can also be obtained from the Mellin transform method, discussed in
Appendix \ref{sec:RKAP1:Mellin_transform}.

\subsubsection{Asymptotic expansions}

Given the expansions (\ref{eq:RKAP1:Z-function-asymptotic}) and 
\[
Z^{\prime}\left(\zeta\right)\sim\frac{1}{\zeta^{2}}\sum_{m=0}^{\infty}\frac{\left(\nicefrac{3}{2}\right)_{m}}{\zeta^{2m}}\quad\left(\zeta_{i}>0\right),
\]
we can insert these expansions in (\ref{eq:RKAP1:DF-deriv-Superstatistics})
and (\ref{eq:RKAP1:kRPDF-D-Superstatistics}) and then readily obtain
the asymptotic expansions 
\begin{multline}
Z_{\kappa}^{(\eta,\zeta,\mu)\prime}\left(\xi\right)\sim\sqrt{\pi\eta}\tilde{D}_{\kappa}^{(\eta,\zeta,\mu)}\frac{\eta}{\xi^{2}}\\
\times\sum_{k=0}\left(\frac{3}{2}\right)_{k}U\left(k+\frac{3}{2},k+\frac{5}{2}-\zeta,\mu\eta\right)\left(\frac{\eta}{\xi^{2}}\right)^{k},\label{eq:RKAP1:kRPDF-deriv-asymptotic}
\end{multline}
and 
\begin{multline}
D_{\kappa}^{(\eta,\zeta,\mu)}\left(\xi\right)\sim-\sqrt{\pi}\tilde{D}_{\kappa}^{(\eta,\zeta,\mu)}\frac{\sqrt{\eta}}{\xi}\\
\times\sum_{k=0}\left(\frac{1}{2}\right)_{k}U\left(k+\frac{1}{2},k+\frac{3}{2}-\zeta,\mu\eta\right)\left(\frac{\eta}{\xi^{2}}\right)^{k},\label{eq:RKAP1:D-function-asymptotic}
\end{multline}
both valid for $\left|\xi\right|\gg \sqrt{\eta}$.

All representations obtained in this section for the functions $Z_{\kappa}^{(\eta,\zeta,\mu)}\left(\xi\right)$ and $Z_{\kappa}^{(\eta,\zeta,\mu)\prime}\left(\xi\right)$ are sufficient for their numerical implementation and for the analytical study of wave propagation and wave-particle interaction in RKD plasmas.

\subsection{Plots of the dispersion function and its derivative}

We will now show some plots of the functions $Z_{\kappa}^{(\eta,\zeta,\mu)}\left(\xi\right)$
and $Z_{\kappa}^{(\eta,\zeta,\mu)\prime}\left(\xi\right)$ for the
model $\eta=\kappa$ and $\zeta=\kappa+1$ as a function of $\xi$\@.
The plots will also contain the limits $\mu\to0$, corresponding to
the $\kappa$PDF $Z_{\kappa}^{\left(1,0\right)}\left(\xi\right)$,
given by (\ref{eq:RKAP1:SKD-kPDF}), and subsequently $\kappa\to\infty$,
which will recover the Fried \& Conte function defined by (\ref{eq:RKAP1:Fried-Conte-DF}).

\begin{figure}
\noindent \includegraphics[width=1\columnwidth]{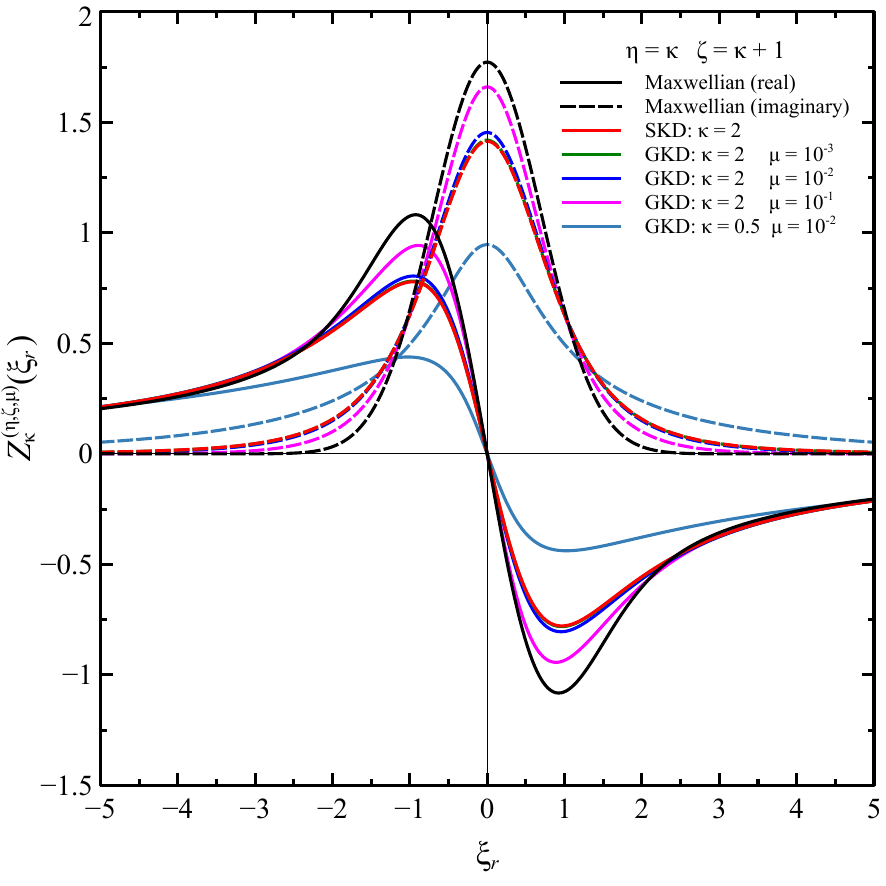}

\noindent \includegraphics[width=1\columnwidth]{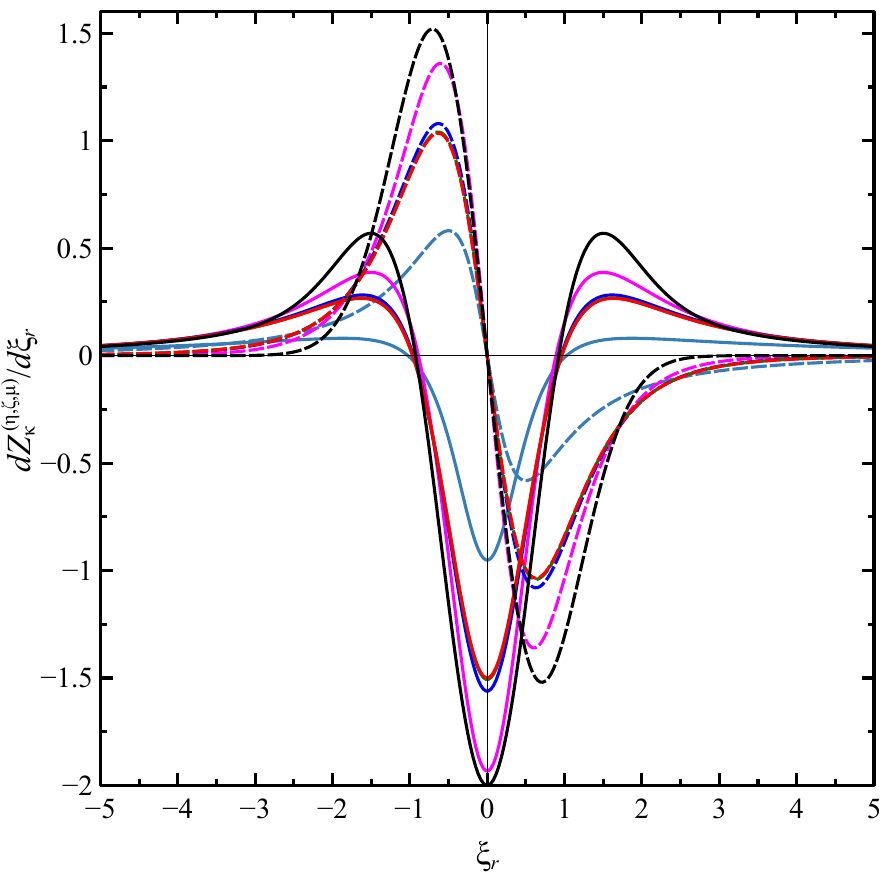}

\caption{Plots of $Z_{\kappa}^{(\eta,\zeta,\mu)}\left(\xi\right)$ (top panel)
and $Z_{\kappa}^{(\eta,\zeta,\mu)\prime}\left(\xi\right)$ (bottom
panel) as functions of $\xi=\xi_{r}$ for the model $\eta=\kappa$
and $\zeta=\kappa+1$\@. \label{fig:RKAP1:Plots-Zrk-Zrkp}}
\end{figure}

Figure \ref{fig:RKAP1:Plots-Zrk-Zrkp} shows some plots of the function
$Z_{\kappa}^{(\eta,\zeta,\mu)}\left(\xi\right)$ and its derivative
as functions of real $\xi$\@. The top panel of Fig. \ref{fig:RKAP1:Plots-Zrk-Zrkp}
contains plots of the Fried \& Conte function as a function of $\xi_{r}$
identified by black curves. The black continuous curve, labelled ``Maxwellian
(real)'', shows the real part of $Z\left(\xi\right)$, whereas the
black dashed line, labelled ``Maxwellian (imaginary)'' shows the
imaginary part. The same convention of real parts shown as continuous
lines and imaginary parts shown as dashed lines is followed by all
plots in both panels. The top panel also contains the plots of the
$\kappa$PDF $Z_{\kappa}^{(1,0)}\left(\xi_{r}\right)$ for the case
$\kappa=2$ drawn by red lines and labelled ``SKD: $\kappa=2$''\@.
Comparison of the red and black curves show the differences between
the dispersion functions of Maxwellian and suprathermal plasmas, in
both real and imaginary parts. The remaining curves in the top panel
are plots of the regularized $\kappa$ dispersion function (R$\kappa$PDF)
for different values of the parameters $\kappa$ and $\mu$, with
all curves identified by labels starting with ``GKD''\@. For all
GKD curves with $\kappa=2$, one can observe how the increase of $\mu$,
from $\mu=10^{-3}$ to $\mu=10^{-1}$, displaces the plots of the
R$\kappa$PDF from the standard suprathermal towards the Maxwellian
case. It is worth mentioning here that the plots of the R$\kappa$PDF
for $\kappa=2$ and $\mu=10^{-3}$ (green curves) are indistinguishable
from the $\kappa$PDF (red lines)\@. Finally, an extreme case with
$\kappa=\nicefrac{1}{2}$ and $\mu=10^{-2}$ is also shown, which
is not possible for dispersion functions from SKDs, since in this
case $\kappa>\nicefrac{1}{2}$.

The bottom panel of figure \ref{fig:RKAP1:Plots-Zrk-Zrkp} contains
the plots of $Z_{\kappa}^{(\eta,\zeta,\mu)\prime}\left(\xi_{r}\right)$\@.
The same conventions for continuous and dashed lines and color codes
used in the top panel apply here. Likewise, the red curves for the
$\kappa$PDF with $\kappa=2$ are not discernible from the green curves
for the R$\kappa$PDF with $\kappa=2$ and $\mu=10^{-3}$\@. In the
same manner, as $\mu$ increases, the R$\kappa$PDF plots approach
the $Z\left(\xi_{r}\right)$ plots.

\begin{figure*}
\begin{minipage}[t]{0.49\textwidth}%
\noindent \includegraphics[width=1\columnwidth]{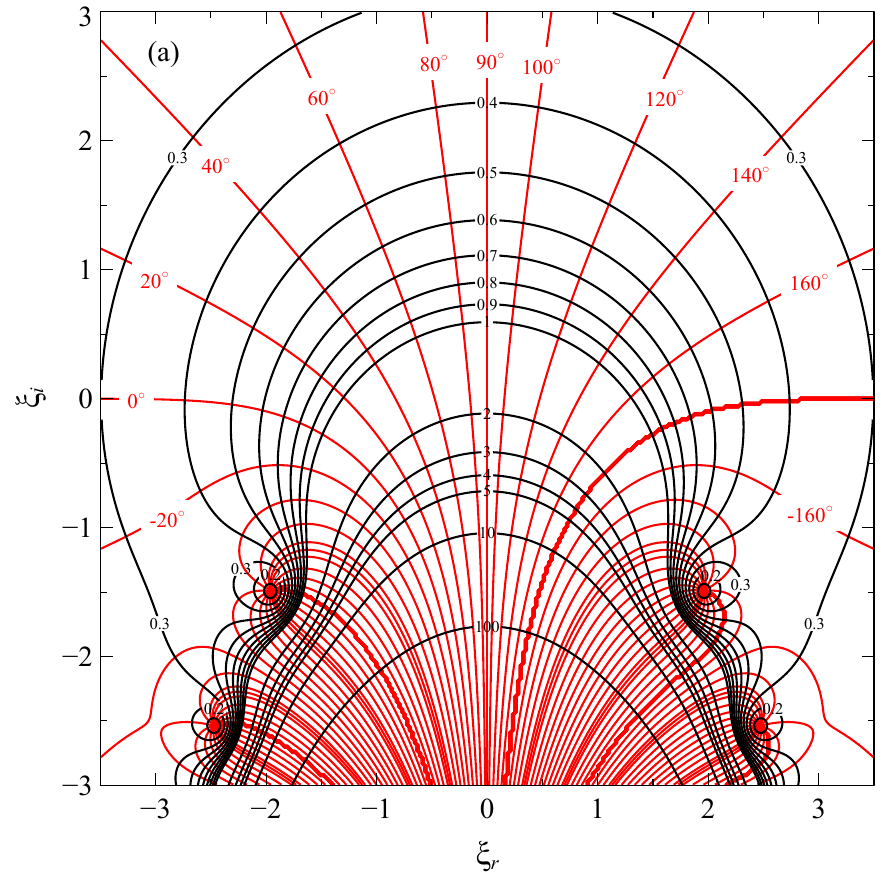}%
\end{minipage}\hfill{}%
\begin{minipage}[t]{0.49\textwidth}%
\noindent \includegraphics[width=1\columnwidth]{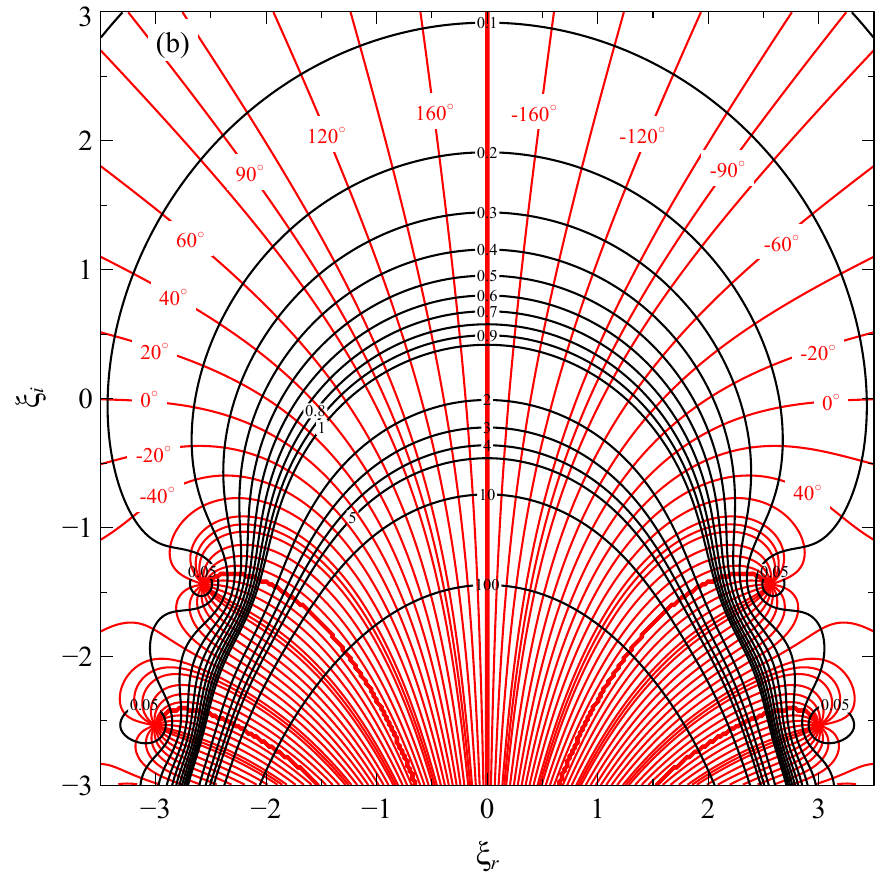}%
\end{minipage}

\begin{minipage}[t]{0.49\textwidth}%
\noindent \includegraphics[width=1\columnwidth]{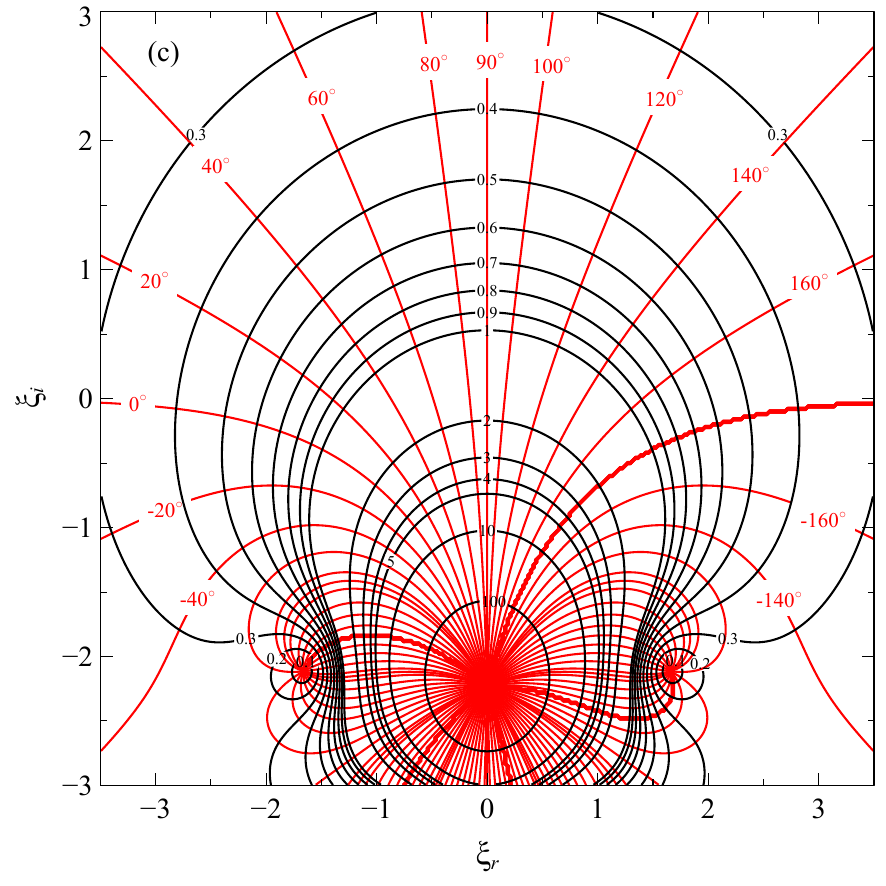}%
\end{minipage}\hfill{}%
\begin{minipage}[t]{0.49\textwidth}%
\noindent \includegraphics[width=1\columnwidth]{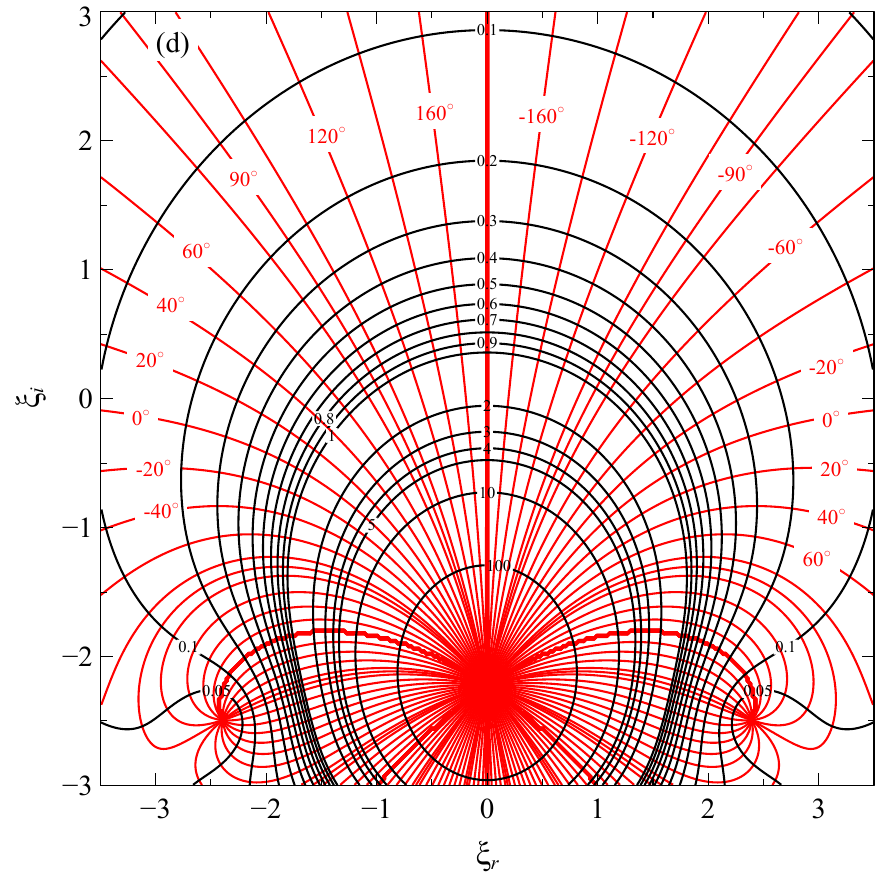}%
\end{minipage}

\caption{Amplitude-phase charts of $Z_{\kappa}^{(\eta,\zeta,\mu)}\left(\xi\right)$
(panels a and c) and $Z_{\kappa}^{(\eta,\zeta,\mu)\prime}\left(\xi\right)$
(panels b and d)\@. In panels (a) and (b): $\kappa=25$ and $\mu=10^{-3}$\@.
In panels (c) and (d): $\kappa=5$ and $\mu=10^{-3}$.\label{fig:RKAP1:Amplitude-phase-charts}}
\end{figure*}

Now, in figure \ref{fig:RKAP1:Amplitude-phase-charts} we include
some amplitude-phase charts of $Z_{\kappa}^{(\eta,\zeta,\mu)}\left(\xi\right)$
(in panels a and c) and $Z_{\kappa}^{(\eta,\zeta,\mu)\prime}\left(\xi\right)$
(in panels b and d) in the complex plane of $\xi=\xi_{r}+i\xi_{i}$
for the same model. In these charts, a complex number $z$ is represented
by its modulus $\left|z\right|$ and phase (or argument) $\theta$,
\emph{i.e.}, $z=\left|z\right|e^{i\theta}$\@. In all panels, the
black contour curves depict the modulus and the red lines show the
phase in degrees $\left(-180^{\circ}<\theta<180^{\circ}\right)$.

In panels (a) and (b) the parameters are $\kappa=25$ and $\mu=10^{-3}$\@.
For such large value of $\kappa$ and small value of $\mu$, the charts
for the R$\kappa$PDF and its derivative are very similar to the respective
plots of $Z\left(\xi\right)$ and $Z^{\prime}\left(\xi\right)$\@.
The most prominent features in these plots are the zeroes of the functions,
which are symmetrically distributed around the imaginary axis. In
the $\xi$-plane region depicted, one notices in panel (a) the roots
of $Z_{\kappa}^{(\eta,\zeta,\mu)}\left(\xi\right)$ at $\xi\approx\pm1.99-1.5i$
and $\xi\approx\pm2.5-2.5i$\@. Likewise, in panel (b) one observes
the roots of $Z_{\kappa}^{(\eta,\zeta,\mu)\prime}\left(\xi\right)$
located at $\xi\approx\pm2.57-1.45i$ and $\xi\approx\pm2.97-2.52i$\@.

On the other hand, in panels (c) and (d) the parameters are $\kappa=5$
and $\mu=10^{-3}$\@. In comparison with the previous plots, some
interesting differences for this suprathermal plasma can be observed.
First of all, only a pair of roots remain for both functions; at $\xi\approx1.66-2.11i$
for $Z_{\kappa}^{(\eta,\zeta,\mu)}\left(\xi\right)$ (panel c) and
at $\xi\approx\pm2.41-2.5i$ for $Z_{\kappa}^{(\eta,\zeta,\mu)\prime}\left(\xi\right)$
(panel d)\@. However, the most remarkable feature now are the singularities
located at $\xi=-i\sqrt{5}$\@. For integer $\kappa$, the singularities
at $\xi=-i\sqrt{\kappa}$ are simple poles, but for noninteger $\kappa$
we have branch points at these points.

\begin{figure*}
\begin{minipage}[t]{0.49\textwidth}%
\noindent \includegraphics[width=1\columnwidth]{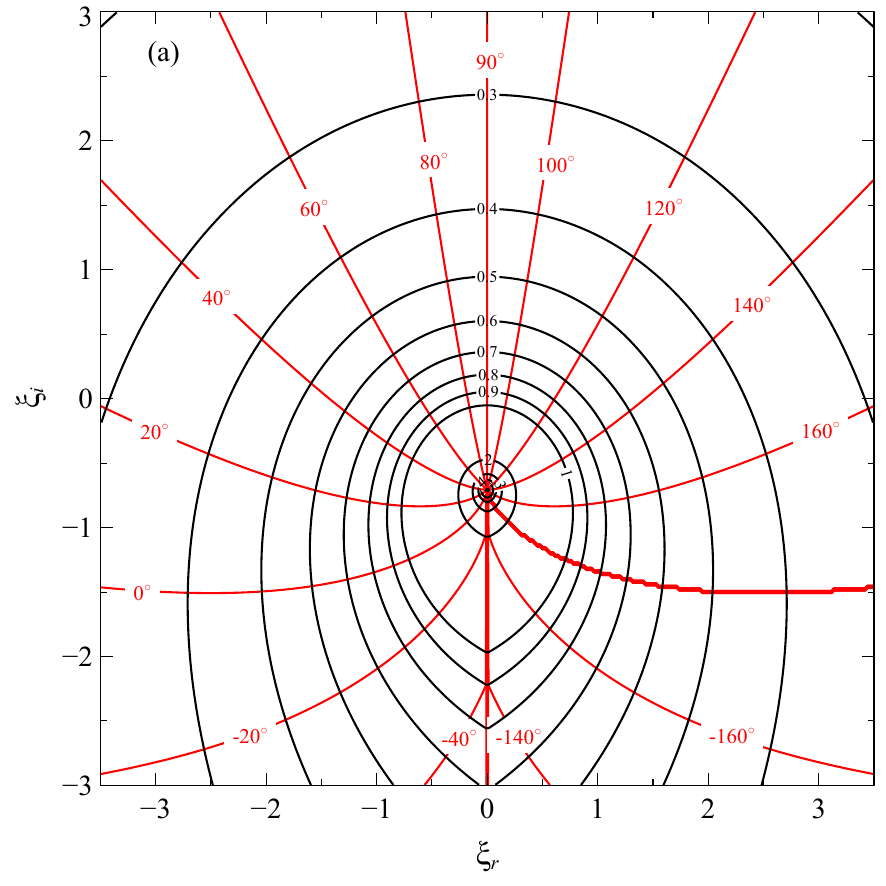}%
\end{minipage}\hfill{}%
\begin{minipage}[t]{0.49\textwidth}%
\noindent \includegraphics[width=1\columnwidth]{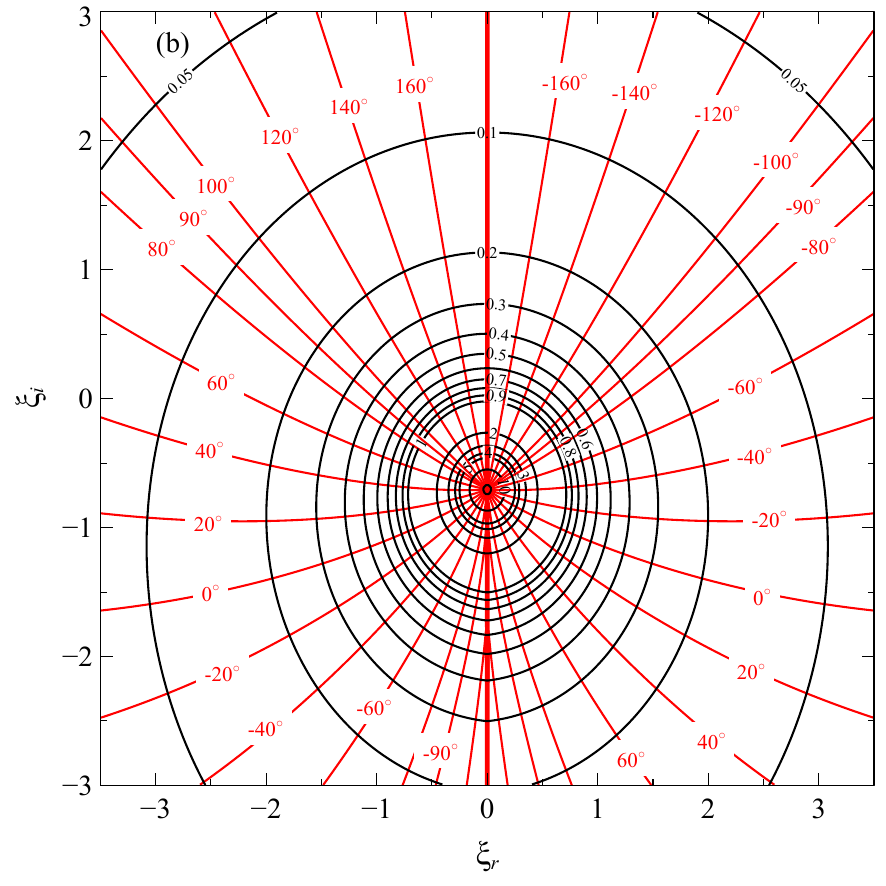}%
\end{minipage}

\caption{Amplitude-phase charts of $Z_{\kappa}^{(\eta,\zeta,\mu)}\left(\xi\right)$
(panel a) and $Z_{\kappa}^{(\eta,\zeta,\mu)\prime}\left(\xi\right)$
(panel b) for the case $\kappa=\nicefrac{1}{2}$ and $\mu=10^{-2}$\@.\label{fig:RKAP1:Amplitude-phase-charts-2}}
\end{figure*}

A last amplitude-phase chart is shown in figure \ref{fig:RKAP1:Amplitude-phase-charts-2},
now for the extreme suprathermal plasma with $\kappa=\nicefrac{1}{2}$
and $\mu=10^{-2}$, which can not exist in RKD plasmas. The plots
are somewhat simpler than the previous ones, because the roots of
the PDF and its derivative have moved away from the depicted region
of the complex $\xi$ plane. However, the branch point at $\xi=-i/\sqrt{2}$
and the associated branch cut along the imaginary axis towards $-i\infty$
is clearly seen.

\section{The dispersion relation and absorption coefficient of Langmuir waves}\label{sec:RKAP1:Langmuir}

A simple application of the formalism developed in this work would
be the analysis of high-frequency longitudinal waves propagating in
a field-free plasma (or in the direction of the ambient magnetic field),
more commonly known as Langmuir waves.

For a general gyrotropic plasma, the dispersion equation for the longitudinal
modes is given by (\ref{eq:RKAP1:DE-electrostatic})\@. In a pure
RKD plasma, the dispersion equation, according to (\ref{eq:RKAP1:DTMPI-e3}),
is
\[
\Lambda\left(k_{\parallel},\omega\right)=\varepsilon_{3,\mathrm{RKD}}=1-\sum_{a}\frac{\omega_{pa}^{2}}{k_{\parallel}^{2}\theta_{a}^{2}}Z_{\kappa}^{(\eta,\zeta,\mu)\prime}\left(\frac{\omega}{k_{\parallel}\theta_{a}}\right)=0.
\]

\begin{subequations}
\label{eq:RKAP1:DE-Langmuir-various}

Since Langmuir waves have high frequency, ususally only the contribution
of the electrons is included. In this case, the dispersion equation
becomes
\begin{equation}
\Lambda\left(k_{\parallel},\omega\right)=1-\frac{\omega_{pe}^{2}}{k_{\parallel}^{2}\theta_{e}^{2}}Z_{\kappa}^{(\eta,\zeta,\mu)\prime}\left(\frac{\omega}{k_{\parallel}\theta_{e}}\right)=0.\label{eq:RKAP1:DE-Langmuir-RKD}
\end{equation}
Some numerical solutions of (\ref{eq:RKAP1:DE-Langmuir-RKD}), obtained
as $\omega=\omega_{r}\left(k_{\parallel}\right)+i\omega_{i}\left(k_{\parallel}\right)$,
will be shown momentarily, along with solutions of the equations in
the consecutive limits $\mu\to0$ and $\kappa\to\infty$, which are,
respectively,
\begin{gather}
\Lambda_{\mathrm{SKD}}\left(k_{\parallel},\omega\right)=1-\frac{\omega_{pe}^{2}}{k_{\parallel}^{2}\theta_{e}^{2}}Z_{\kappa}^{\left(1,0\right)\prime}\left(\frac{\omega}{k_{\parallel}\theta_{e}}\right)=0,\label{eq:RKAP1:DE-Langmuir-SKD}\\
\Lambda_{M}\left(k_{\parallel},\omega\right)=1-\frac{\omega_{pe}^{2}}{k_{\parallel}^{2}\theta_{e}^{2}}Z^{\prime}\left(\frac{\omega}{k_{\parallel}\theta_{e}}\right)=0.\label{eq:RKAP1:DE-Langmuir-M}
\end{gather}
These equations will be solved for the model $\eta=\kappa$, $\zeta=\kappa+1$
and $\theta_{e}=\mathrm{const.}$ (independent of $\kappa$)\@. In
this case, the parameter $\theta_{e}$ is simply the thermal speed
of electrons, $\theta_{e}=\sqrt{2T_{e}/m_{e}}$.
\end{subequations}

The numerical solutions of (\ref{eq:RKAP1:DE-Langmuir-various}a-c)
will be compared with analytical solutions obtained from the weak
absorption approximation, valid when $\left|\omega_{i}\right|\ll\omega_{r}$
and $\left|\omega_{i}\partial\Lambda_{i}/\partial\omega_{r}\right|\ll\left|\Lambda_{r}\left(\omega_{r}\right)\right|$\@.
In this case, writing $\Lambda\left(\omega_{r}\right)=\Lambda_{r}\left(\omega_{r}\right)+i\Lambda_{i}\left(\omega_{r}\right)$,
expression (\ref{eq:RKAP1:Z_k-AC}) shows that
\begin{align*}
\Lambda_{r}\left(k_{\parallel},\omega_{r}\right) & =1-\frac{\omega_{pe}^{2}}{k_{\parallel}^{2}\theta_{e}^{2}}Z_{\kappa,NC}^{(\eta,\zeta,\mu)\prime}\left(\frac{\omega_{r}}{k_{\parallel}\theta_{e}}\right)\\
\Lambda_{i}\left(k_{\parallel},\omega_{r}\right) & =-\pi\frac{\omega_{pe}^{2}}{k_{\parallel}^{2}\theta_{e}^{2}}Z_{\kappa,C}^{(\eta,\zeta,\mu)\prime}\left(\frac{\omega_{r}}{k_{\parallel}\theta_{e}}\right).
\end{align*}
Then, the dispersion relation $\omega_{r}=\omega_{r}\left(k_{\parallel}\right)$
will be given by the solution of $\Lambda_{r}\left(k_{\parallel},\omega_{r}\right)=0$,
whereas the absorption rate coefficient $\omega_{i}\left(k_{\parallel}\right)$
is given approximately by
\[
\omega_{i}=-\frac{\Lambda_{i}\left(k_{\parallel},\omega_{r}\right)}{\partial\Lambda_{r}/\partial\omega_{r}}.
\]

Langmuir waves are considered fast waves, \emph{i.e.}, $\omega_{r}/k_{\parallel}\gg\theta_{e}$\@.
Hence, on $\Lambda_{r}\left(k_{\parallel},\omega_{r}\right)$ we can
employ the first few terms of the asymptotic expansion (\ref{eq:RKAP1:kRPDF-deriv-asymptotic}),
and write
\begin{multline*}
Z_{\kappa,NC}^{(\eta,\zeta,\mu)\prime}\left(\frac{\omega_{r}}{k_{\parallel}\theta_{e}}\right)\approx\sqrt{\pi\eta}\tilde{D}_{\kappa}^{(\eta,\zeta,\mu)}\eta\frac{k_{\parallel}^{2}\theta_{e}^{2}}{\omega_{r}^{2}}\\
\times\left[U\left(\frac{3}{2},\frac{5}{2}-\zeta,\mu\eta\right)+\frac{3}{2}U\left(\frac{5}{2},\frac{7}{2}-\zeta,\mu\eta\right)\eta\frac{k_{\parallel}^{2}\theta_{e}^{2}}{\omega_{r}^{2}}\right].
\end{multline*}
This leads to the expression
\[
\Lambda_{r}\left(k_{\parallel},\omega_{r}\right)\approx1-\frac{\omega_{pe}^{2}}{\omega_{r}^{2}}\left(\mathcal{A}_{1}+\frac{3}{2}\mathcal{A}_{2}\frac{k_{\parallel}^{2}\theta_{e}^{2}}{\omega_{r}^{2}}\right),
\]
where
\begin{align*}
\mathcal{A}_{1} & =\sqrt{\pi\eta}B_{\kappa}^{(\eta,\zeta,\mu)}U\left(\frac{3}{2},\frac{5}{2}-\zeta,\mu\eta\right)=1\\
\mathcal{A}_{2} & =\sqrt{\pi\eta}\eta B_{\kappa}^{(\eta,\zeta,\mu)}U\left(\frac{5}{2},\frac{7}{2}-\zeta,\mu\eta\right)\\
 & =\eta\frac{U\left(\nicefrac{5}{2},\nicefrac{7}{2}-\zeta,\mu\eta\right)}{U\left(\nicefrac{3}{2},\nicefrac{5}{2}-\zeta,\mu\eta\right)}\stackrel{(\ref{eq:RKAP1:Kummer-transformation})}{=}\frac{1}{\mu}\frac{U\left(\zeta,\zeta-\nicefrac{3}{2},\mu\eta\right)}{U\left(\zeta,\zeta-\nicefrac{1}{2},\mu\eta\right)}.
\end{align*}

\begin{subequations}
\label{eq:RKAP1:DE-Langmuir-RKD-WA}
Therefore, there results the R$\kappa$-Bohm-Gross dispersion relation
\begin{gather}
\omega_{RBG}^{(\eta,\zeta,\mu)}\left(k_{\parallel}\right)\approx\omega_{pe}\sqrt{1+\frac{3}{2}\mathcal{A}_{2}\frac{k_{\parallel}^{2}\theta_{e}^{2}}{\omega_{pe}^{2}}}.
\end{gather}

Now, according to the derivative of (\ref{eq:RKAP1:Z_k,C^(e,z,m)}),
\[
\frac{\Lambda_{i}\left(k_{\parallel},\omega_{r}\right)}{2\pi\tilde{D}_{\kappa}^{(\eta,\zeta,\mu)}}=\frac{\omega_{pe}^{2}\omega_{r}}{k_{\parallel}^{3}\theta_{e}^{3}}\left(1+\frac{1}{\eta}\frac{\omega_{r}^{2}}{k_{\parallel}^{2}\theta_{e}^{2}}\right)^{-\zeta}\exp\left(-\frac{\mu\omega_{r}^{2}}{k_{\parallel}^{2}\theta_{e}^{2}}\right),
\]
and approximating $\partial\Lambda_{r}/\partial\omega_{r}\approx2/\omega_{pe}$,
we obtain the absorption coefficient in a RKD plasma
\begin{multline}
\frac{\omega_{i,RBG}^{(\eta,\zeta,\mu)}}{\omega_{RBG}^{(\eta,\zeta,\mu)}}\approx-\sqrt{\pi}\mathcal{A}_{3}\left(\frac{1}{\sqrt{\eta}}\frac{\omega_{pe}}{k_{\parallel}\theta_{e}}\right)^{3}\\
\times\left(1+\frac{1}{\eta}\frac{\omega_{r}^{2}}{k_{\parallel}^{2}\theta_{e}^{2}}\right)^{-\zeta}\exp\left(-\frac{\mu\omega_{r}^{2}}{k_{\parallel}^{2}\theta_{e}^{2}}\right),
\end{multline}
where $\mathcal{A}_{3}=\left[\left(\mu\eta\right)^{\zeta-3/2}U\left(\zeta,\zeta-\nicefrac{1}{2},\mu\eta\right)\right]^{-1}$.
\end{subequations}

\begin{subequations}
\label{eq:RKAP1:Langmuir-WA-model-limits}

Expressions (\ref{eq:RKAP1:DE-Langmuir-RKD-WA}a,b) are generic. For
the model $\eta=\kappa$, $\zeta=\kappa+1$ and $\theta_{e}=\mathrm{const.}$,
these expressions, along with the corresponding ones derived from
the limits $\mu\to0$ and $\kappa\to\infty$, become, respectively,
the pairs of expressions
\begin{gather}
\begin{aligned}\omega_{RBG}^{(\kappa,\kappa+1,\mu)}\left(k_{\parallel}\right) & =\omega_{pe}\sqrt{1+\frac{3}{2}\mathcal{A}_{2}\frac{k_{\parallel}^{2}\theta_{e}^{2}}{\omega_{pe}^{2}}}\\
\frac{\omega_{i,RBG}^{(\kappa,\kappa+1,\mu)}}{\omega_{RBG}^{(\kappa,\kappa+1,\mu)}} & =-\sqrt{\pi}\mathcal{A}_{3}\left(\frac{1}{\sqrt{\kappa}}\frac{\omega_{pe}}{k_{\parallel}\theta_{e}}\right)^{3}\\
 & \times\left(1+\frac{1}{\kappa}\frac{\omega_{r}^{2}}{k_{\parallel}^{2}\theta_{e}^{2}}\right)^{-\left(\kappa+1\right)}\exp\left(-\frac{\mu\omega_{r}^{2}}{k_{\parallel}^{2}\theta_{e}^{2}}\right),
\end{aligned}
\label{eq:RKAP1:LWAML-RKD}\\
\begin{aligned}\omega_{SKD}\left(k_{\parallel}\right) & =\omega_{pe}\sqrt{1+\frac{3}{2}\frac{\kappa}{\kappa-\nicefrac{3}{2}}\frac{k_{\parallel}^{2}\theta_{e}^{2}}{\omega_{pe}^{2}}}\\
\frac{\omega_{i,SKD}}{\omega_{SKD}} & =-\sqrt{\pi}\frac{\Gamma\left(\kappa+1\right)}{\Gamma\left(\kappa-\nicefrac{1}{2}\right)}\\
 & \times\left(\frac{1}{\sqrt{\kappa}}\frac{\omega_{pe}}{k_{\parallel}\theta_{e}}\right)^{3}\left(1+\frac{1}{\kappa}\frac{\omega_{r}^{2}}{k_{\parallel}^{2}\theta_{e}^{2}}\right)^{-\left(\kappa+1\right)},
\end{aligned}
\label{eq:RKAP1:LWAML-SKD}\\
\begin{aligned}\omega_{BG}\left(k_{\parallel}\right) & =\omega_{pe}\sqrt{1+\frac{3}{2}\frac{k_{\parallel}^{2}\theta_{e}^{2}}{\omega_{pe}^{2}}}\\
\frac{\omega_{i}}{\omega_{BG}} & =-\sqrt{\pi}\left(\frac{\omega_{pe}}{k_{\parallel}\theta_{e}}\right)^{3}e^{-\omega_{r}^{2}/k_{\parallel}^{2}\theta_{e}^{2}},
\end{aligned}
\label{eq:RKAP1:LWAML-M}
\end{gather}
where now
\begin{align*}
\mathcal{A}_{2} & =\frac{1}{\mu}\frac{U\left(\kappa+1,\kappa-\nicefrac{1}{2},\mu\kappa\right)}{U\left(\kappa+1,\kappa+\nicefrac{1}{2},\mu\kappa\right)}, & \mathcal{A}_{3}^{-1} & =U\left(\frac{3}{2},\frac{3}{2}-\kappa,\mu\kappa\right).
\end{align*}

Expressions (\ref{eq:RKAP1:LWAML-RKD}) approximately describe the
propagation and absorption of Langmuir waves in a RKD plasma for the
adopted model, whereas expressions (\ref{eq:RKAP1:LWAML-SKD}) describe
the Langmuir waves in a SKD plasma for the same model. Those waves
only can exist when $\kappa>\nicefrac{3}{2}$ and other models will
render different expressions for either $\omega_{r}\left(k_{\parallel}\right)$
or $\omega_{i}\left(k_{\parallel}\right)$\@. Finally, expressions
(\ref{eq:RKAP1:LWAML-M}) are the usual Bohm-Gross dispersion relation
and the absorption coefficient in a Maxwellian plasma.
\end{subequations}

\begin{figure*}
\begin{minipage}[t]{0.49\textwidth}%
\noindent \includegraphics[width=1\columnwidth]{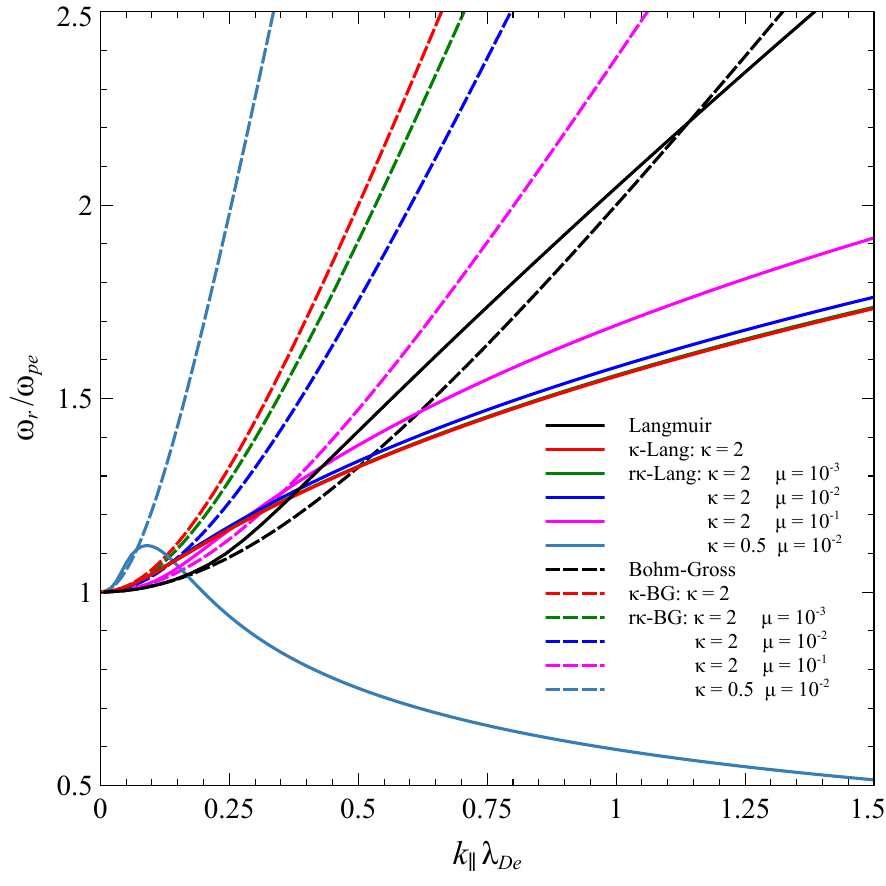}%
\end{minipage}\hfill{}%
\begin{minipage}[t]{0.49\textwidth}%
\noindent \includegraphics[width=1\columnwidth]{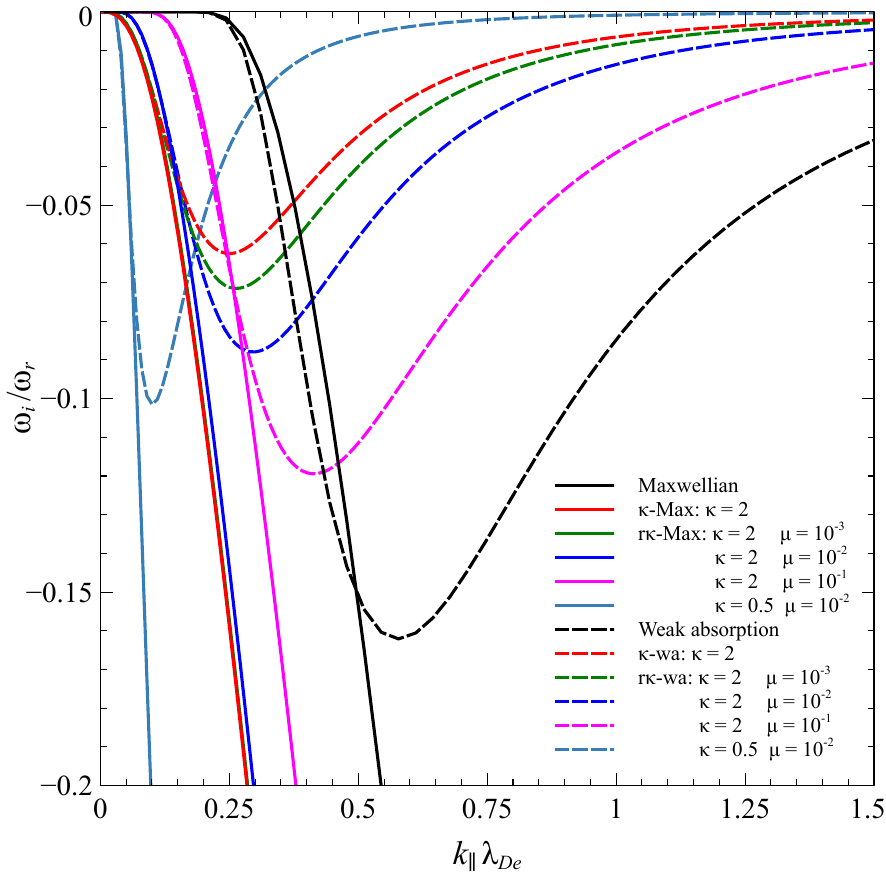}%
\end{minipage}

\caption{Plots of the real (left panel) and imaginary (right panel) parts of
the solutions of the dispersion equation (\ref{eq:RKAP1:DE-Langmuir-RKD})
and of the weak-absorption expressions (\ref{eq:RKAP1:DE-Langmuir-RKD-WA}a,b).\label{fig:RKAP1:Langmuir_waves}}
\end{figure*}

Figure \ref{fig:RKAP1:Langmuir_waves} shows plots of the real and
imaginary parts of the numerical solutions of the dispersion equations
(\ref{eq:RKAP1:DE-Langmuir-various}a-c), as well as of the approximate
expressions (\ref{eq:RKAP1:Langmuir-WA-model-limits}a-c), as functions
of $k_{\parallel}\lambda_{De}$, where $\lambda_{De}=\sqrt{T_{e}/4\pi n_{e}e^{2}}=\theta_{e}/\sqrt{2}\omega_{pe}$
is the Debye length for electrons.

In the left panel we display the dispersion relations. The label ``Langmuir''
identifies the numerical solution of the dispersion equation (\ref{eq:RKAP1:DE-Langmuir-M})
for Langmuir waves in a Maxwellian plasma, whereas ``$\kappa$-Lang''
identifies the corresponding dispersion relation in a SKD plasma with
$\kappa=2$ for the adopted model, as a solution of (\ref{eq:RKAP1:DE-Langmuir-SKD})\@.
The remaining continuous curves, labelled with ``r$\kappa$-Lang'',
correspond to the dispersion relations in a RKD plasma with $\kappa=2$
and several values of $\mu$, as solutions of (\ref{eq:RKAP1:DE-Langmuir-RKD})\@.
We also plot the extreme case of the dispersion relation for Langmuir
waves in a RKD plasma with $\kappa=\nicefrac{1}{2}$ and $\mu=10^{-2}$,
which can not exist in a pure SKD plasma. The black dashed curve corresponds
to the Bohm-Gross dispersion relation and those with labels ``$\kappa$-BG''
and ``r$\kappa$-BG'' are the corresponding dispersions for SKD
and RKD plasmas, given by (\ref{eq:RKAP1:Langmuir-WA-model-limits}a-c).

A striking feature in these plots is the overall divergence between
a given numerical (``exact'') dispersion relation and the corresponding
approximation. What really happens is that the region of validity
of the approximate expressions in a suprathermal plasma diminishes
with the $\kappa$ index. Observing the black curves, the Bohm-Gross
dispersion relation gives a fairly good description until it crosses
with the numerical solution, at $k_{\parallel}\lambda_{De}\approx1.15$\@.
For smaller wavenumbers, $\omega_{BG}\left(k_{\parallel}\right)$
is slightly smaller than $\omega_{L}\left(k_{\parallel}\right)$ (the
numerical solution), but for larger wavenumbers $\omega_{BG}\left(k_{\parallel}\right)$
crosses $\omega_{L}\left(k_{\parallel}\right)$ and rapidly diverges,
becoming less and less accurate. The same happens with all the other
pairs of curves, but since the SKD and RKD plasmas are highly superthermal
(with $\kappa=2$), the crossings occur at much smaller wavenumbers,
at $k_{\parallel}\lambda_{De}\lesssim0.1$\@. Consequently, the approximate
dispersion relations become inaccurate much sooner.

Another remarkable result is given by the extreme case $\kappa=\nicefrac{1}{2}$\@.
Such small value of the $\kappa$ index means that there is a substantially
large number of electrons in the mid-range suprathermal tail of the
VDF\@. Langmuir waves propagating in a system so far from thermal
equilibrium display the opposite tendency seen in a thermal plasma:
the wave frequency rapidly begins to reduce with the wavenumber, becoming
even smaller than the plasma frequency. These results indicate that
the propagation of Langmuir waves in suprathermal plasma has distinctly
different properties than in a thermal plasma.

The same convention was largely adopted for the imaginary parts (right
panel)\@. The label ``Maxwellian'' identifies the absorption rate
coefficient $\left(\omega_{i}\right)$ for a Maxwellian plasma, obtained
from the numerical solution of the dispersion equation (\ref{eq:RKAP1:DE-Langmuir-M}),
and ``Weak absorption'' (wa) identifies the corresponding approximation
(\ref{eq:RKAP1:LWAML-M})\@. Accordingly, ``$\kappa$-Max'', ``r$\kappa$-Max'',
``$\kappa$-wa'' and ``r$\kappa$-wa'' identify the absorption
coefficients for SKD and RKD plasmas.

We observe that the region of validity of the approximations is now
even smaller than for the dispersion relations. For instance, the
absorption of waves in a thermal plasma is reasonably well-described
by the approximation up to $k_{\parallel}\lambda_{De}\approx0.5$\@.
Once again, the region of validity becomes smaller as the $\kappa$
index reduces. We also observe that Langmuir waves propagating in
the extreme RKD plasma (with $\kappa=\nicefrac{1}{2}$) are strongly
absorbed by the suprathermal electrons. In this case, the region of
validity of the approximation extends only up to $k_{\parallel}\lambda_{De}\lesssim0.1$.

\section{Summary}\label{conclusions}
In this paper, we continued the development of the theoretical formalism
involved with the proposition of the regularized kappa velocity distribution
function as a model for space plasmas that are in a nonequilibrium
quasi-stationary state. 

Starting from the generalized form of the regularized kappa distribution,
two new functions were introduced, the associated plasma dispersion
function and its derivative, which describe wave propagation and wave-particle
interactions (in the linear regime) for the particular case of parallel
propagation relative to the ambient magnetic field.

Several properties, series and integral representations for the new
functions were obtained, which are useful for the theoretical description
of wave-particle interactions taking place in a suprathermal plasma,
and also for the numerical implementation of the dispersion functions,
with the subsequent derivation of the dispersion relations and wave
damping/growth rates of the normal modes that are sustained by the
plasma.
Some plots of the new dispersion functions were shown, both along the real line of the argument and on the complex plane.

The formalism developed here was applied to the study of propagation and absorption of Langmuir waves in a pure RKD plasma.

In future publications, the formalism that was developed in this work
will be applied to several other
relevant situations involving wave propagation
and interaction with suprathermal plasmas, particularly for the cases
when temperature-anisotropy instabilities arise. The mathematical
formalism can also be used to validate recent computational codes
that were proposed to simulate suprathermal plasmas.

\begin{acknowledgments}
RG acknowledges support provided by Conselho Nacional de Desenvolvimento
Científico e Tecnológico (CNPq), Grant. no. 313330/2021-2. We are also grateful for support from the Deutsche Forschungsgemeinschaft (DFG) within the framework of the lead agency ('weave') project FI 706/31-1 and the DFG project 334/16-1. 
\end{acknowledgments}

\appendix

\section{Series representation of $D_{\kappa}^{(\eta,\zeta,\mu)}\left(\xi\right)$
from Mellin transform\label{sec:RKAP1:Mellin_transform}}

Methods based on the Mellin transform\citep{ParisKaminski01} were
already successfully employed by Refs. \onlinecite{GaelzerZiebell14/12,GaelzerZiebell16/02,Gaelzer+16/06}
to derive several properties of the standard kappa dispersion function
$Z_{\kappa}^{\left(\alpha,\beta\right)}\left(\xi\right)$\@. We will
now show that the same methods can also be useful for the more complicated
case of the regularized $\kappa-$distribution.

We will employ the Mellin transform method to derive a series representation
for the function $D_{\kappa}^{(\eta,\zeta,\mu)}\left(\xi\right)$,
defined by (\ref{eq:RKAP1:kRPDF-D2-D})\@. We start by writing the
definition as 
\[
D_{\kappa}^{(\eta,\zeta,\mu)}\left(\xi\right)=\tilde{D}_{\kappa}^{(\eta,\zeta,\mu)}\frac{\xi}{\sqrt{\eta}}\int_{0}^{\infty}dt\frac{t^{-1/2}e^{-\mu\eta t}}{t-\xi^{2}/\eta}\left(1+t\right)^{-\zeta}.
\]

Introducing the identity\citep{Prudnikov90v3}
\[
\frac{1}{t-\xi^{2}/\eta}=-\frac{\eta}{\xi^{2}}\frac{1}{2\pi i}\int_{L}\Gamma\left(-s\right)\Gamma\left(1+s\right)\left(-\frac{\eta}{\xi^{2}}t\right)^{s}ds,
\]
where the integration is carried along the adequate contour of the
inverse Mellin transform,\citep{ParisKaminski01} we can switch integrations
and write 
\begin{multline*}
D_{\kappa}^{(\eta,\zeta,\mu)}\left(\xi\right)=-\tilde{D}_{\kappa}^{(\eta,\zeta,\mu)}\frac{\sqrt{\eta}}{\xi}\times\\
\frac{1}{2\pi i}\int_{L}ds\,\Gamma\left(-s\right)\Gamma\left(1+s\right)\left(-\frac{\eta}{\xi^{2}}\right)^{s}\\
\times\int_{0}^{\infty}dt\,t^{s-1/2}\left(1+t\right)^{-\zeta}e^{-\mu\eta t}.
\end{multline*}

The last integration is carried out using (\ref{eq:RKAP1:Tricomi-Integ_rep-1}),
\begin{multline*}
\int_{0}^{\infty}dt\,t^{s-1/2}(1+t)^{-\zeta}\mathrm{e}^{-\mu\eta t}\\
=\Gamma\left(s+\frac{1}{2}\right)U\left(s+\frac{1}{2},s+\frac{3}{2}-\zeta,\mu\eta\right),
\end{multline*}
which is valid for $\mathrm{Re}\left(s+\nicefrac{1}{2}\right)>0$.

If we now write 
\begin{align*}
\left(-\right)^{s} & =e^{i\pi s}=\cos\pi s+i\sin\pi s\\
 & =\frac{\pi}{\Gamma\left(\nicefrac{1}{2}-s\right)\Gamma\left(\nicefrac{1}{2}+s\right)}-\frac{i\pi}{\Gamma\left(-s\right)\Gamma\left(1+s\right)},
\end{align*}
where we have used the identity \citep{AskeyRoy-NIST10}
\[
\Gamma\left(z\right)\Gamma\left(1-z\right)=\frac{\pi}{\sin\left(\pi z\right)},
\]
we then obtain 
\begin{equation}
D_{\kappa}^{(\eta,\zeta,\mu)}\left(\xi\right)=-\pi\tilde{D}_{\kappa}^{(\eta,\zeta,\mu)}\frac{\sqrt{\eta}}{\xi}\left(\mu\eta\right)^{\zeta-1/2}\left(I_{s}^{(1)}-iI_{s}^{(2)}\right),\label{eq:RKAP1:D-I1-I2}
\end{equation}
where 
\begin{align*}
I_{s}^{(1)} & =\frac{1}{2\pi i}\int_{L}ds\,\frac{\Gamma\left(-s\right)\Gamma\left(1+s\right)}{\Gamma\left(\nicefrac{1}{2}-s\right)}\frac{U\left(\zeta,\zeta+\nicefrac{1}{2}-s,\mu\eta\right)}{\left(\mu\xi^{2}\right)^{s}}\\
I_{s}^{(2)} & =\frac{1}{2\pi i}\int_{L}ds\,\Gamma\left(\frac{1}{2}+s\right)\frac{U\left(\zeta,\zeta+\nicefrac{1}{2}-s,\mu\eta\right)}{\left(\mu\xi^{2}\right)^{s}},
\end{align*}
after employing again the Kummer transformation (\ref{eq:RKAP1:Kummer-transformation}).

Looking at $I_{s}^{(1)}$ and $I_{s}^{(2)}$, we observe that the
gamma functions in the integrand of $I_{s}^{(1)}$ have simple poles
at $s=-\ell-1$ $\left(\ell=0,1,2,\dots\right)$, whereas in $I_{s}^{(2)}$
the poles occur at $s=-\ell-\nicefrac{1}{2}$ $\left(\ell=0,1,2,\dots\right)$\@.
At all these poles, and in sufficiently small neighborhoods, the Tricomi
functions are regular. Therefore, we can employ the residue theorem
for both integrals.

For $I_{s}^{(1)}$, the integrand is 
\[
f_{1}\left(s\right)=\frac{\Gamma\left(-s\right)\Gamma\left(1+s\right)}{\Gamma\left(\nicefrac{1}{2}-s\right)}\left(\mu\xi^{2}\right)^{-s}U\left(\zeta,\zeta+\frac{1}{2}-s,\mu\eta\right).
\]
The residues at $s=s_{\ell}=-\ell-1$ $\left(\ell=0,1,2,\dots\right)$
are\footnote{See, \emph{e.g.}, identity \url{http://functions.wolfram.com/06.05.06.0009.01}.}
\[
\left.\mathrm{res}f_{1}\left(s\right)\right|_{s=s_{\ell}}=\left(-\right)^{\ell}\frac{\left(\mu\xi^{2}\right)^{\ell+1}}{\Gamma\left(\ell+\nicefrac{3}{2}\right)}U\left(\zeta,\zeta+\ell+\frac{3}{2},\mu\eta\right).
\]
Hence, 
\[
I_{s}^{(1)}=\frac{2}{\sqrt{\pi}}\left(\mu\xi^{2}\right)\sum_{\ell=0}^{\infty}\frac{\left(-\mu\xi^{2}\right)^{\ell}}{\left(\nicefrac{3}{2}\right)_{\ell}}U\left(\zeta,\zeta+\ell+\frac{3}{2},\mu\eta\right).
\]

Now for $I_{s}^{(2)}$, the integrand is 
\[
f_{2}\left(s\right)=\Gamma\left(\frac{1}{2}+s\right)\left(\mu\xi^{2}\right)^{-s}U\left(\zeta,\zeta+\frac{1}{2}-s,\mu\eta\right),
\]
for which simple poles occur at $s=s_{\ell}=-\ell-\nicefrac{1}{2}$
$\left(\ell=0,1,\dots\right)$\@. The residues of $f_{2}\left(s\right)$
at $s=s_{\ell}$ are
\[
\left.\mathrm{res}f_{2}\left(s\right)\right|_{s=s_{\ell}}=\frac{\left(-\right)^{\ell}}{\ell!}\left(\mu\xi^{2}\right)^{\ell+\nicefrac{1}{2}}U\left(\zeta,\zeta+\ell+1,\mu\eta\right).
\]
Therefore, 
\[
I_{s}^{(2)}=\left(\mu\xi^{2}\right)^{1/2}\sum_{\ell=0}^{\infty}\frac{\left(-\right)^{\ell}}{\ell!}\left(\mu\xi^{2}\right)^{\ell}U\left(\zeta,\zeta+\ell+1,\mu\eta\right).
\]

However,\citep{Daalhuis-NIST10a},
\[
U\left(a,a+n+1,z\right)=z^{-a}\sum_{k=0}^{n}\binom{n}{k}\left(a\right)_{k}z^{-k}.
\]
Hence, 
\[
I_{s}^{(2)}=\left(\mu\xi^{2}\right)^{1/2}\left(\mu\eta\right)^{-\zeta}S_{2}\left(\zeta;\mu\eta,-\frac{\xi^{2}}{\eta}\right),
\]
where 
\[
S_{2}\left(\zeta;x,y\right)=\sum_{\ell=0}^{\infty}\sum_{k=0}^{\ell}\binom{\ell}{k}\frac{\left(\zeta\right)_{\ell-k}}{\ell!}x^{k}y^{\ell}.
\]

But, given the following formula,\citep{SrivastavaManocha84}
\[
\sum_{\ell=0}^{\infty}\sum_{k=0}^{\ell}B\left(k,\ell\right)=\sum_{\ell=0}^{\infty}\sum_{k=0}^{\infty}B\left(k,\ell+k\right),
\]
we can write 
\[
S_{2}\left(\zeta;x,y\right)=\left(\sum_{\ell=0}^{\infty}\left(\zeta\right)_{\ell}\frac{y^{\ell}}{\ell!}\right)\left(\sum_{k=0}^{\infty}\frac{\left(xy\right)^{k}}{k!}\right)=\pFq 10\left({\zeta\atop -};y\right)e^{xy},
\]
where $\pFq 10\left(\zeta,-;y\right)$ is a generalized hypergeometric
function.\footnote{For a short introduction to hypergeometric functions, see appendix
B of Ref. \onlinecite{GaelzerZiebell16/02}.}

However, 
\[
\pFq 10\left({a\atop -};z\right)=\left(1-z\right)^{-a}\Longrightarrow S_{2}\left(\zeta;x,y\right)=\left(1-y\right)^{-\zeta}e^{xy}.
\]
Therefore, we obtain
\[
I_{s}^{(2)}=\left(\mu\xi^{2}\right)^{1/2}\left(\mu\eta\right)^{-\zeta}\left(1+\frac{\xi^{2}}{\eta}\right)^{-\zeta}e^{-\mu\xi^{2}}.
\]

Returning to (\ref{eq:RKAP1:D-I1-I2}), we finally obtain
\begin{multline*}
D_{\kappa}^{(\eta,\zeta,\mu)}\left(\xi\right)=-\sqrt{\pi}\tilde{D}_{\kappa}^{(\eta,\zeta,\mu)}\\
\times\left[2\frac{\xi}{\sqrt{\eta}}\sum_{\ell=0}^{\infty}\frac{y_{\ell}\left(\zeta,\zeta+\nicefrac{3}{2};\mu\eta\right)}{\left(\nicefrac{3}{2}\right)_{\ell}}\left(-\frac{\xi^{2}}{\eta}\right)^{\ell}\right.\\
\left.-\sqrt{\pi}i\left(1+\frac{\xi^{2}}{\eta}\right)^{-\zeta}e^{-\mu\xi^{2}}\right],
\end{multline*}
which is exactly (\ref{eq:RKAP1:RKD-series-1}).


%

\end{document}